\documentclass{article}
\usepackage[in]{fullpage}
\usepackage{amsmath,amssymb,amsthm,mathtools,mathrsfs,bm}
\usepackage{helvet}
\usepackage{courier}
\usepackage[hyphens]{url}
\usepackage{graphicx}
\usepackage[square,numbers]{natbib}
\usepackage{algorithm}
\usepackage{algcompatible}

\usepackage{gensymb}
\usepackage{color}
\usepackage{xcolor}
\usepackage{mathtools}
\usepackage{verbatim}
\usepackage{amsmath,amssymb,amsfonts}
\usepackage{algorithm}
\usepackage{graphicx}
\usepackage{textcomp}
\usepackage{xcolor}
\usepackage{amsthm}
\usepackage{multirow}
\usepackage{cleveref}
\usepackage[switch]{lineno}
\allowdisplaybreaks
\usepackage[]{algpseudocode}

\title{Online Task Assignment with Controllable Processing Time}
\author{
Ruoyu Wu\thanks{School of Computer Science, The University of Sydney, ruwu6940@uni.sydney.edu.au} \and
Wei Bao\thanks{School of Computer Science, The University of Sydney, \{wei.bao, liming.ge\}@sydney.edu.au} \and
Liming Ge\textsuperscript{$\dag$}
}
\date{}

\usepackage{amsmath,amsthm,amssymb,mathtools,bm}
\usepackage{multirow}
\usepackage{comment}
\usepackage{subcaption}
\usepackage{refcount}

\usepackage{todonotes}
\newtheorem{theorem}{Theorem}

\newtheorem{lemma}{Lemma}

\usepackage{enumerate}
\usepackage{enumitem}
\usepackage{booktabs}

\newcommand*{\defeq}{\stackrel{\text{def}}{=}}

\begin{document}

\maketitle

\begin{abstract}
We study a new online assignment problem, called the \textit{Online Task Assignment with Controllable Processing Time}. In a bipartite graph, a set of online vertices (tasks) should be assigned to a set of offline vertices (machines) under the known adversarial distribution (KAD) assumption. We are the first to study controllable processing time in this scenario: There are  multiple processing levels for each task and higher level brings larger utility but also larger processing delay. A machine can reject an assignment at the cost of a rejection penalty, taken from a pre-determined rejection budget. Different processing levels cause different penalties.
We propose the Online Machine and Level Assignment  (OMLA) Algorithm to simultaneously assign an offline machine and a processing level to each online task. We prove that OMLA achieves $1/2$-competitive ratio if each machine has unlimited rejection budget and $\Delta/(3\Delta-1)$-competitive ratio if each machine has an initial rejection budget up to $\Delta$.
Interestingly, the competitive ratios do not change under different settings on the controllable processing time and we can conclude that OMLA is ``insensitive'' to the controllable processing time. 
\end{abstract}

\section{Introduction}\label{sec:1}
In this paper, we study an online task assignment problem with controllable processing time. In this problem, we have a set of online vertices (tasks) and a set of offline vertices (machines). Online tasks arrive sequentially and each can be processed by a machine, but each task can only be processed by a subset of machines~\cite{RN49,RN51}. 
We focus on the controllable processing time: Each task has multiple levels of processing time~\cite{RN81,RN68}. If a task is processed with a higher level, it obtains a higher reward, but needs to wait for a longer time. 
Each machine can only process one task at one time, and can only take another after the previous one is finished.
The machines do not always accept task assignment, and they lose an amount of rejection budget every time they reject. The higher processing level of the rejected assignment is, the larger amount of the budget is taken. When a machine runs out of its rejection budget, it will be removed from the system immediately and permanently. In this paper, we consider the online tasks arriving under \textit{known adversarial distributions} (KAD)~\cite{RN49,RN61}. The arrival probability of each task  at each time is known ahead. The goal is to maximise the expected reward 
without the knowledge of future task arrivals (online setting). 

Although a number of works have studied the online assignment problem, this paper is the first to consider controllable processing time. This is motivated by real-world scenarios in various fields, such as:



\textbf{1. Task offloading in edge computing}~\cite{RN63}. With the help of edge computing, end user devices can offload computing intensive tasks to edge computers, especially the ones processing machine learning (ML) models. Each user (task) has only a set of edge computers (machines)  near them, so the tasks from this user can only be processed by these computers. When we assign an edge computer to a task, we can further control the processing time by implementing different ML models. A better processing quality requires a model with longer processing time, but a lightweight  model (e.g., a pruned and sparsified model) finishes sooner, with lower accuracy. 

\textbf{2. Ride-sharing with tolls}~\cite{RN62}. Ride-sharing system assigns passenger requests (tasks) to available drivers (machines). Each request has an origin and can only be processed by drivers near this area. For each ride, we can choose to go through toll roads (high cost) for a shorter processing time, or to avoid toll for less cost. 

\textbf{3. Translation service}~\cite{RN85}. Customers place orders (tasks) to the language service agencies (machines), and the agency can provide different degrees of service (e.g., one-off translation, translation and proofreading, etc.). Each translation request has a target language and can only be processed by translators with a certification of this language. The agency can choose to assign more time for a translation task, resulting in a higher processing quality; but the agency can also assign a shorter processing time to a task, so that more customer requests can be processed.


In many situations, because the arrival of tasks can only be known when they arrive, online algorithms are required. We are motivated to design such an online algorithm that can maximize the worst ratio against the offline optimal performance (competitive ratio). In short, the online assignment problem studied in this paper has the following features:\\
\textbf{A. Known adversarial distribution (KAD):} the probability of the arrival of each task at each time is known in advance.\\
\textbf{B. Reusable machines:} the machine returns to the system after completing a task; the processing delay is drawn from known distributions.\\
\textbf{C. Controllable processing time:} each task can be processed with different levels; a higher processing level generates a higher reward but the expected delay is also higher.\\
\textbf{D. Budgeted machine rejections:} when a machine is assigned, it can reject the assignment with a penalty; rejecting a higher processing level task will cause higher penalty. 
When the budget runs out, the machine is permanently removed from the system.

The main contribution of this paper is designing an Online Machine and Level Assignment (OMLA) Algorithm for the above problem, especially with multiple processing levels. We  prove that our algorithm achieves a $1/2$-competitive ratio when every machine has an infinite rejection budget, and a $\Delta/(3\Delta-1)$-competitive ratio when each machine has a finite rejection budget, where $\Delta$ is the largest budget of machines at the beginning. The conclusion shows that regardless of the limited rejection budgets, the competitive ratio does not depend on the  processing levels, indicating that OMLA is insensitive to controllable processing time.

Controllable processing time makes the problem studied in the paper more realistic but also introduces substantially more challenges to the online algorithm design and competitive analysis. Controllable processing time expands the searching space. Since each processing level causes different rewards, delays, and rejection budgets. The algorithm should balance these dimensions as a result of coupled objective and constraints. To tackle this challenge, in our online algorithm design, we first use the joint probabilities of choosing a machine and  a level as decision variables to formulate an offline linear programming (LP). The optimal solution to the LP is then leveraged to calculate the activation value and the baseline value for each machine and level. These two values will determine our decision on the machine and level when we make decisions online. To bound the competitive ratio, we introduce a reference system where each task with $L$ levels are reconstructed as $L$ tasks with a single level. Then the performance of the reference system is employed as an intermediate value to bound the competitive ratio. Mathematical derivations demonstrate that multiple processing levels do not worsen the competitive ratio because the reference system uniformly bounds different processing levels, and thus the competitive ratio is insensitive to the controllable processing time.

\section{Related Work}\label{sec:2}

One category of works related to this paper is \textit{Online Bipartite Matching}, where the system needs to assign offline machines to online tasks to maximize the utility~\cite{RN66}. One subcategory of works focuses on the adversary arrival order~\cite{RN64}, and another subcategory assumes known adversarial distribution (KAD)~\cite{RN47,RN65} or known identical independent distributions (KIID)~\cite{RN68}, where task arrival  follows known distributions. Motivated by real-world scenarios, \cite{RN84} and \cite{RN49} studied the case that machines are reusable, and \cite{RN52,RN53,RN62} studied the case that machines can reject task assignment. \cite{RN51} studied both reusable machines and rejections. Other topics studied in this field include fairness for task assignment~\cite{RN50,RN54}, multi-unit demand (a task may need multiple machines to process)~\cite{RN46,RN55}, and multi-capacity agent (a machine can process multiple tasks)~\cite{RN80,RN56}. However, there is no existing work considering controllable processing time in the online bipartite matching
problem. A majority part of \cite{RN51} can be regarded as a special case of our work when controllable processing time is not considered.  It gives a $1/2$-competitive algorithm when each offline machine can reject unlimited times, and a $\Delta/(3\Delta-1)$-competitive algorithm when each machine can reject no more than $\Delta$ times. Interestingly, our proposed algorithm also gives the same competitive ratios, but with substantially more complicated designs and analyses. To this end, a key conclusion derived in our paper is that the competitive ratio is ``insensitive'' to the processing levels.  Please note that another work~\cite{RN79} studied controllable reward and different arrival probabilities, where the assignment impacts reward and arrival probabilities, which is different from controllable processing time
in nature. Online bipartite matching is leveraged to solve many real-world problems other than machine allocation, such as ride-sharing~\cite{RN47,RN49,RN50}, crowd-sourcing~\cite{RN46,RN99,RN79} and AdWords~\cite{RN78}. Still none of the existing work considered controllable processing time.

Another category of works related to this paper is \textit{Controllable Processing Time}. Controllable processing time is studied in the context of scheduling. We can reduce the processing time of a job at a cost of reduced processing reward~\cite{RN68,RN69}. \cite{RN70}, \cite{RN71} and \cite{RN72} study single machine scheduling. \cite{RN75} and \cite{RN77} study multiple parallel machine scheduling. \cite{RN76} employs bipartite matching to analyze multiple parallel machine scheduling. The above works focus on the offline scheduling problem. 
\cite{RN73} and \cite{RN74} study online scheduling with controllable processing time. \cite{RN73} focuses on single machine scheduling and \cite{RN74} focuses on the flow shop scheduling.
Controllable processing time is also analyzed in the context of stochastic lot-sizing problem. We can compress the production time with extra cost, so that a better performance of planning can be obtained~\cite{RN57,RN82,RN59}. These works optimize the performance in an offline manner. To the best of our knowledge, no existing work considered controllable processing time for the bipartite matching problem.

\section{Model}\label{sec:3}

We present a formal description of our problem in this section.  We have a bipartite graph $G=(U,V;E)$, where $U$ is the set of machines and $V$ is the set of repeatable tasks.\footnote{Please note that  $v\in V$ is actually indicating a type of tasks, which  arrive repeatably at different time. For presentation convenience and following the convention, $v$ is also called ``a task'' throughout this paper.} $E$ is the set of edges indicating if a task $v\in V$ can be processed by a machine $u\in U$. For each task, we have $L$ processing levels $\mathcal{L}=\{1,...,L\}$, indicating the $L$ quality levels. If task $v$ is processed by machine $u$ ($(u,v)\in E$) with processing level $l$, it will generate a reward of $r_{u,v,l}$. Without loss of generality (WLOG), we have $ r_{u,v,l}<r_{u,v,l'}$ when $l<l'$ (larger processing level gives larger reward). The system runs on a finite time horizon $T\in \mathbb{N}^+$. Each processing level $l\in\mathcal{L}$ causes a random processing delay $d_l$, which presents the occupation time to process a task with level $l$. In other words, if a machine  starts to process a task with level $l$, it becomes unavailable to any other tasks until time $t+d_l$. $d_l$ is drawn from a known distribution $D_l$. We have $\mathbb{E}[d_l]<\mathbb{E}[d_{l'}]$ when $l<l'$ (larger processing level requires longer processing time). For the convenience, we denote the set of edges connected to task $v$ by $E_v$ for all task $v$ ($E_v=\{(u,v)|(u,v)\in E\}$), and similarly $E_u$ is set of edges connected to machine $u$.


At each time $t$, task $v$ may arrive with probability $p_{v,t}$. With a probability of $1-\sum_{v\in V}p_{v,t}$, no task arrives at $t$. The set of probability distributions $\{\{p_{v,t}\}_{v\in V}\}_{t\in [T]}$ is known to us in advance (time variant but independent across time).  When a task $v$ arrives, we immediately and irrevocably either assign one machine which is a neighbor to $v$ and is available, or discard $v$. Besides, when we assign task $v$ to machine $u$, we also specify a processing level $l$. When receiving the assignment of task $v$, machine $u$ has two possible actions: with probability $q_e$, $u$ accepts the assignment; with probability $1-q_e$, $u$ rejects the assignment ($e=(u,v)$). Suppose this assignment is specified with processing level $l$, these two actions have two different results. If $u$ accepts this assignment, it immediately gets a reward $r_{u,v,l}$ and becomes unavailable for a random period $d_l$.  A machine $u$ has a limited rejection budget (initialized as $\Delta_u$).
If $u$ rejects this assignment, a rejection-penalty $\theta_l$ is introduced. We assume that $\Delta_u$ and $\theta_l$ are integers. This penalty is taken from the remaining rejection budget of $u$, denoted by $\delta_u$. For convenience, we denote $\theta=\max_{l\in\mathcal{L}}\theta_l$. 
When a machine $u$ runs out of its remaining budget ($\delta_u\leq0$), it is removed from the system immediately and permanently. If a machine $u$ is removed from the system, it receives no more task assignments. 

Each machine has a positive initial budget ($\Delta_u>0$).
Please note we allow $\Delta_u=\infty$ to indicate unlimited rejection. We also allow $\theta_l=1$ to indicate homogeneous rejection penalty (to limit the number of rejections). 

\subsection{Solution Overview}\label{sec:3.1}
Our objective is to maximize the sum reward. We focus on the online setting: we only know the arrival of a task when it arrives. We know the distribution of task arrival in advance and the distribution of occupation time (KAD).

We first construct a linear programming (LP) \texttt{Off} to get an optimal solution $\mathbf{x}^*$ and an upper bound of the offline optimal value, which is referred to as LP(\texttt{Off}). The optimal solution $\mathbf{x}^*$ is then employed to construct our online algorithm. In the meanwhile, the upper bound of the offline optimal value LP(\texttt{Off}) will be set as a benchmark to evaluate the online algorithm, so that we then evaluate the competitive ratio between the performance of online algorithm and the offline optimal value. 

\subsection{Offline Optimal Value and Competitive Ratio}\label{sec:3.2}

We consider the offline optimization version of the original problem as the benchmark and define the competitive ratio. In the offline setting, the full task sequence $I$ is known in advance. However, we do not know whether a machine will accept or reject an assignment until it happens. We only have the probability of  acceptance $q_e$.
Given a full task sequence $I$, if an offline algorithm maximizes the expected reward, it is an offline optimal algorithm for $I$. This maximized expected reward for $I$ is denoted by OPT($I$). 
The expected OPT($I$) on every sequence $I$ is $\mathbb{E}_{I\sim\mathcal{I}}[\text{OPT}(I)]$, which is referred to as the \textit{offline optimal value}.


{We say that an online algorithm ALG is $\alpha$-competitive if the expected reward obtained by ALG is at least $\alpha$ times the offline optimal value. That is, if $\mathbb{E}_{I\sim\mathcal{I}}[\text{ALG}(I)]\geq\alpha\mathbb{E}_{I\sim\mathcal{I}}[\text{OPT}(I)]$ for any $\mathcal{I}$.}




\subsection{Linear Programming}\label{sec:3.3}

It is not straightforward to quantify the offline optimal value
$\mathbb{E}_{I\sim\mathcal{I}}[\text{OPT}(I)]$. In what follows, 
we construct an offline LP to get the upper bound of the offline optimal value.  

\begin{align}
    \max_{\substack{x_{e,l,t},\\ \forall e\in E, \\l\in \mathcal{L}, \\t\in[T]}} &\ \displaystyle{\sum_{t\in[T]}\sum_{e\in E}q_e\sum_{l\in\mathcal{L}}}r_{u,v,l} x_{e,l,t}\nonumber\\
    \text{s.t. } & \displaystyle{\sum_{t'<t}\sum_{e\in E_u}q_e\sum_{l\in\mathcal{L}}}x_{e,l,t'}\text{Pr}\{d_l\geq t-t'+1\}  +\displaystyle{\sum_{e\in E_u}q_e\sum_{l\in\mathcal{L}}}x_{e,l,t}\leq1,\ \ (\forall u\in U,t\in [T]),\label{eqn:offst1}\\
    &\displaystyle{\sum_{t\in[T]}\sum_{e\in E_u}\sum_{l\in\mathcal{L}}}x_{e,l,t}\big[\theta q_e\text{Pr}\{d_l>T-t\}+(1-q_e)\theta_l\big]\leq\Delta_u+\theta-1,(\forall u\in U),\label{eqn:offst2}\\
    & 0\leq\displaystyle{\sum_{e\in E_v}\sum_{l\in\mathcal{L}}}x_{e,l,t}\leq p_{v,t},\ \ (\forall v\in V, t\in[T]),\label{eqn:offst4}\\
    & 0\leq\displaystyle{\sum_{l\in\mathcal{L}}}x_{e,l,t}\leq p_{v,t},(\forall v\in V,e\in E_v,t\in[T]),\label{eqn:offst5}\\
    & 0\leq\displaystyle{\sum_{e\in E_u}\sum_{l\in\mathcal{L}}}x_{e,l,t}\leq1,\ \ (\forall u\in U,t\in[T]),\label{eqn:offst6}
\end{align}
This LP is referred to as \texttt{Off}. The optimal solution to \texttt{Off} is $\mathbf{x}^*\defeq \{x_{e,l,t}^*\}$, and the optimal value for the objective function of \texttt{Off} is referred to as LP(\texttt{Off}). LP(\texttt{Off}) is the upper bound for $\mathbb{E}_{I\sim\mathcal{I}}[\text{OPT}(I)]$ (as shown in Lemma below). In addition, $\mathbf{x}^*$ is to be employed in the online algorithm.


In the following Lemma, we show that LP(\texttt{Off}) is a valid upper bound for the offline optimal value $\mathbb{E}_{I\sim\mathcal{I}}[\text{OPT}(I)]$.


\begin{lemma}
\label{lm:1}
 (LP(\texttt{Off}) Upper Bound) LP(\texttt{Off}) $\geq\mathbb{E}_{I\sim \mathcal{I}}[\text{OPT}(I)]$
\end{lemma}

\begin{proof}
Proofs of all Lemmas and Theorems are in the Appendix.
\end{proof}


\subsection{Online Machine and Level Assignment (OMLA) Algorithm }\label{sec:3.4}

In this section, we use the optimal solution to \texttt{Off} to construct our OMLA algorithm.

\textbf{Design Overview.} In the online algorithm, we first decide the probability that we choose machine-level pair $(u,l)$ when task $v$ arrives at time $t$. Then we decide whether or not to assign machine $u$ to task $v$ with processing level $l$, by comparing the different expected rewards (of machine $u$) brought by different decisions. We present our online algorithm (OMLA) in Algorithm \ref{alg:online} and we discuss them line by line.

\begin{algorithm}[t!]
\caption{Online Machine and Level Assignment (OMLA) Algorithm}\label{alg:online}
\textbf{Input:} $U$, $V$, $E$, $\{Q^\delta_{e,l,t}\}$, $\{R^\delta_{u,t}\}$, $\mathbf{x}^*$
    \begin{algorithmic}[1]

    \For{all $t\gets1$ to $T$}
        \If{no task arrives} 
            \State skip
        \Else\ ($v$ arrives)
            \State choose pair $(u,l)$ with probability $x^*_{e,l,t}/p_{v,t}$\label{alg2:choose}
            \If{$u$ is not occupied, $u$ has a positive remaining budget $\delta_u>0$ and $Q^\delta_{e,l,t}\geq R^\delta_{u,t+1}$}\label{alg2:compare}
                \State we assign $(u,l)$ to $v$\label{alg1:9}
                \If{$u$ accepts} 
                    \State draw $d_l$ from $D_l$, $u$ gets occupied for $d_l$\label{alg1:11}
                \Else
                    \State $\delta_u\gets \delta_u-\theta_l$\label{alg1:13}
                \EndIf
            \EndIf
        \EndIf
    \EndFor
        
    \end{algorithmic}
\end{algorithm}


\begin{algorithm}[t]
\caption{Calculation of Activation and Baseline Values}\label{alg:QR}
\textbf{Input:} $U$, $V$, $E$, $\mathcal{L}$, $T$, $\{\theta_l\}$, $\{q_e\}$, $\{r_{u,v,l}\}$, $\{D_l\}, \{\Delta_u\}$
\begin{algorithmic}[1]
\State Solve LP(\texttt{Off}) to obtain $\mathbf{x}^*$
\State $\Delta\gets\max\Delta_u$
\For{all $(\delta,u,v,l)$ that $(u,v)\in E, \delta>0$ and $l\in\mathcal{L}$} \label{QR:3}
        \State $Q^\delta_{e,l,T}\gets q_er_{u,v,l}$ \label{QR:4}
\EndFor
\For{all $(\delta,u)$ that $\delta>0$ and $u\in U$}
    \State $a\gets0$ \label{QR:7}
    \For{all $v$ that $(u,v)\in E_u$}
        \State $b\gets0$
        \For{all $l\in\mathcal{L}$}
            \State $b\gets b+x^*_{e,l,T}q_er_{u,v,l}$
        \EndFor
        \State $a\gets a+b$
    \EndFor
    \State $R^\delta_{u,T}\gets a$
\EndFor \label{QR:15}

        \For{$t\gets T-1$ to $1$}
    \For{$(\delta,u)$ that $\delta>0$ and $u\in U$}
        \For{$(v,l)$ that $(u,v)\in E_u$ and $l\in\mathcal{L}$}
            \State $a\gets0$
            
                \If{$\delta\geq\theta_l$} \label{QR:19}
                    \State $b\gets R^{\delta-\theta_l}_{u,t+1}$\label{QR:20}
                \Else\ 
                     $b\gets0$\label{QR:21}
                \EndIf
                \For{$d\gets1$ to $T-t+1$} \label{QR:23}
                    \State $a\gets a+R^\delta_{u,t+d}$Pr$\{d_l=d\}$ \label{QR:24}
                \EndFor\label{QR:25}
                \State $Q^\delta_{e,l,t}\gets q_e(r_{u,v,l}+a)+(1-q_e)b$\label{QR:26}          
    \EndFor
    \State Calculate $R^\delta_{u,t}$ by (\ref{eqn:defR}) \label{QR:R}
\EndFor
\EndFor
    \end{algorithmic}
    \textbf{Output:} $\{Q^\delta_{e,l,t}\}$, $\{R^\delta_{u,t}\}$, $\mathbf{x}^*$
\end{algorithm}

\textbf{OMLA.} Upon the arrival of task $v$, we first choose a machine $u$ and a processing level $l$ with probability $x^*_{e,l,t}/p_{v,t}$ (Line \ref{alg2:choose} in Algorithm \ref{alg:online}). Suppose $u$ has a remaining rejection budget of $\delta$ at $t$. If $u$ has run out of the rejection budget ($\delta\leq0$) or is occupied by a previous task, we do not assign $u$ or assign any other machine to $v$ (Line \ref{alg2:compare} in Algorithm \ref{alg:online}). 
Then, we define the \emph{baseline value} (R-value) of $u$ at $t$ as the expected sum reward of $u$ at and after $t$ without knowing the arrival at $t$, which is denoted by $R_{u,t}^\delta$; we define the \emph{activation value} (Q-value) of $u$ at $t$ as the expected sum reward of $u$ at and after $t$ if $u$ is assigned to $v$ with processing level $l$, which is denoted by $Q_{e,l,t}^\delta$. More details on the derivations of baseline values and activation values will be given shortly. Baseline value and activation value will be compared to make a decision. We compare activation value at $t$ and the baseline value at $t+1$, to decide if an active action at $t$ (making the assignment) is beneficial. 
If the baseline value of $u$ at $t+1$ is larger, we do not assign $u$ and discard $v$; Otherwise, if the activation value is larger at $t$, we assign $u$ to $v$ with $l$ (Line \ref{alg1:9} in Algorithm \ref{alg:online}). When $u$ accepts this assignment, it becomes occupied for a random time $d_l$ (Line \ref{alg1:11} in Algorithm \ref{alg:online}). When $u$ rejects this assignment, a rejection penalty $\theta_l$ is taken from $u$'s rejection budget $\delta$ (Line \ref{alg1:13} in Algorithm \ref{alg:online}).

\textbf{Calculation of  Activation Values and Baseline Values.} Because we focus on the KAD model, we can calculate each activation value and baseline value in advance (before we execute Algorithm~\ref{alg:online}). The calculation is  presented in Algorithm \ref{alg:QR}.  When $u$ has a positive remaining budget ($\delta>0$), the activation value $Q_{e,l,t}^\delta$ consists of two parts: \textcircled{1} With probability $q_e$ ($e=(u,v)$), $u$ accepts the assignment and immediately gets a reward $r_{u,v,l}$ (\ref{eqn:defQ}). After a random occupation time $d_l$, $u$ finishes this task, and its baseline value becomes $R_{u,t+d_l}^\delta$ at $t+d_l$ (Lines \ref{QR:23}--\ref{QR:25} in Algorithm \ref{alg:QR}); \textcircled{2} With probability $1-q_e$, $u$ rejects the assignment and takes a rejection penalty $\theta_l$ on its remaining budget $\delta$, and its baseline value becomes $R^{\delta-\theta_l}_{u,t+1}$ at $t+1$ (Lines \ref{QR:19}--\ref{QR:21}  in Algorithm \ref{alg:QR}). By combining the above two parts, we can calculate the activation value $Q_{e,l,t}^\delta$ as follows
\begin{linenomath}
\begin{align}
Q_{e,l,t}^\delta=&q_e(r_{u,v,l}+\displaystyle{\sum_{d'\in[T-t]}}\text{Pr}\{d_l=d'\}R_{u,t+d'}^\delta)+(1-q_e)R_{u,t+1}^{\delta-\theta_l},\ \ (\delta>0,t\in[T]).\label{eqn:defQ}
\end{align}
\end{linenomath}

Formula (\ref{eqn:defQ}) is calculated in Line \ref{QR:26}  in Algorithm \ref{alg:QR}.
When $u$ has run out of the rejection budget ($\delta\leq0$), it is removed from the market. We set $Q_{e,l,t}^\delta=0$ if $\delta\leq0$ or $t>T$ as boundary values. Because we choose the higher one between the activation value at $t$ and the baseline value at $t+1$, we have the expected reward of the chosen $u$ as $\max\{Q_{e,l,t}^\delta,R_{u,t+1}^\delta\}$. Since the probability that $v$ arrives at $t$ and $(u,l)$ is chosen is $x^*_{e,l,t}$, we can calculate each baseline value $R_{u,t}^\delta$ (Line \ref{QR:R} in Algorithm \ref{alg:QR}) by
\begin{linenomath}
\begin{align}
    R_{u,t}^\delta=&\displaystyle{\sum_{e\in E_u}\sum_{l\in\mathcal{L}}}x^*_{e,l,t}\max\{Q_{e,l,t}^\delta,R_{u,t+1}^{\delta}\}+(1-\displaystyle{\sum_{e\in E_u}\sum_{l\in\mathcal{L}}}x^*_{e,l,t})R_{u,t+1}^{\delta},(\delta>0,t\in[T]).\label{eqn:defR}
\end{align}
\end{linenomath}
We set $R_{u,t}^\delta=0$ if $\delta\leq0$ or $t>T$ as boundary values.

In order to execute our online algorithm, we need to calculate each $Q_{e,l,t}^\delta$ and $R_{u,t}^\delta$ in advance. This can be done with the initial condition at $T$ (Lines \ref{QR:3}--\ref{QR:15} in Algorithm \ref{alg:QR})
\begin{equation}
    \begin{cases}
        Q_{e,l,T}^\delta=q_er_{u,v,l},&\quad\forall e\in E,l\in\mathcal{L},\delta>0,\\
        R_{u,T}^\delta=\displaystyle{\sum_{e\in E_u}\sum_{l\in\mathcal{L}}}x^*_{e,l,T}q_er_{u,v,l},&\quad\forall u\in U,\delta>0.
    \end{cases}\label{eqn:11}
\end{equation}
We present the algorithm to calculate each $Q_{e,l,t}^\delta$ and $R_{u,t}^\delta$ in Algorithm \ref{alg:QR}. 



\section{Competitive Ratio Analysis}\label{sec:4}

\begin{figure}[h!]{}
     \centering
         \includegraphics[width=0.95\textwidth]{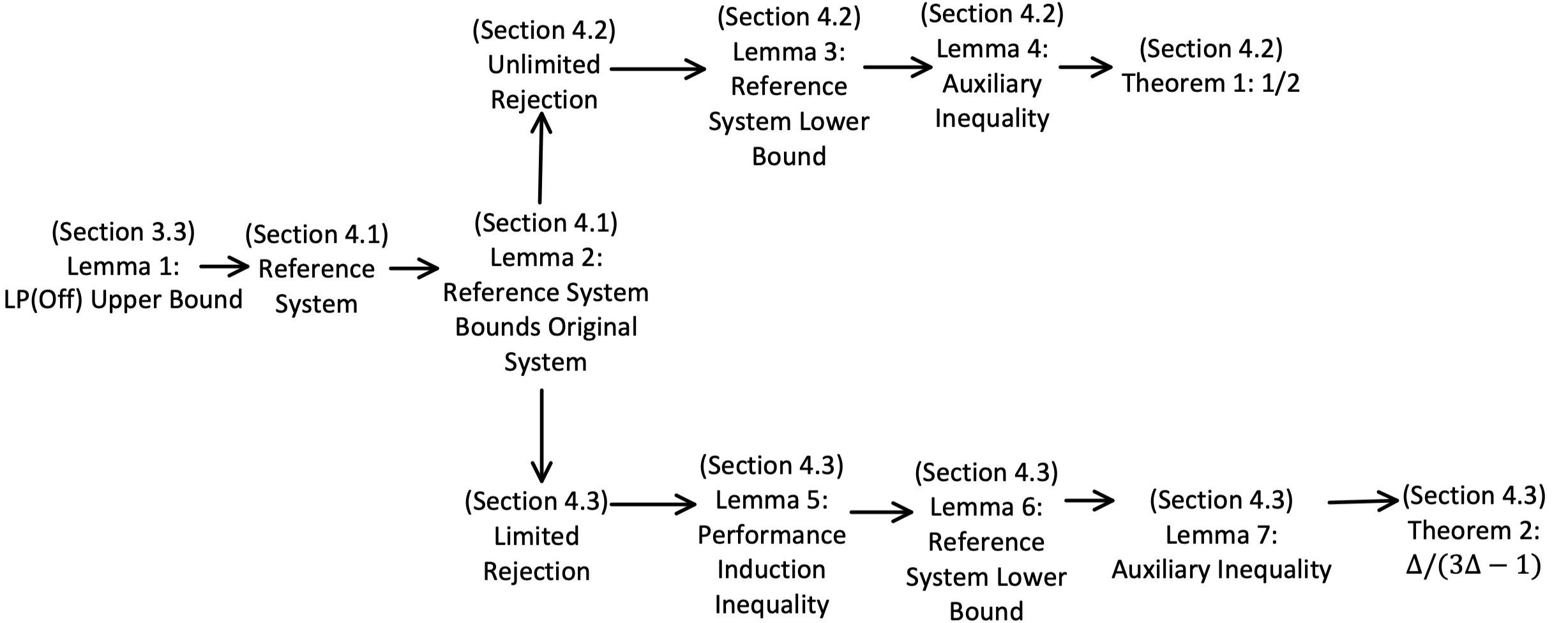}
         
         \caption{Sketch of Analysis.}
         \label{fig:sketch}
\end{figure}

 In this section, we analyze the competitive ratio of our online algorithm. The sketch of our analysis is shown in Figure \ref{fig:sketch}. We have already derived  Lemma~\ref{lm:1}, where we find an upper bound LP(\texttt{Off}) of the offline optimal value. Then we first introduce our reference system, which provides a lower bound of the original system (Lemma \ref{lm:22}). In this reference system, each task with $L$ levels are reconstructed as $L$ tasks with a single level. After Lemma \ref{lm:22}, we analyze the competitive ratio in two branches separately: \textcircled{1} each machine $u$ has an infinite initial rejection budget $\Delta_u=\infty$ (the unlimited rejection case); \textcircled{2} each machine $u$ has a finite rejection budget $\Delta_u<\infty$. For the unlimited rejection case, we first find a lower bound for the reference system (Lemma \ref{lm:3}). Then we construct the \textit{auxiliary inequality} for the unlimited rejection case (Lemma \ref{lm:4}) by Lemmas \ref{lm:1}--\ref{lm:3}. Then by Lemma \ref{lm:4}, we prove that OMLA is $1/2$-competitive for the unlimited rejection case (Theorem \ref{tr:1}). For the limited rejection case, we first find the \textit{performance induction inequality} of the reference system (Lemma \ref{lm:5}). With this inequality, we find a lower bound of the reference system (Lemma \ref{lm:6}). Then we construct the auxiliary inequality for the limited rejection case (Lemma \ref{lm:7}) by Lemmas \ref{lm:1}, \ref{lm:22}, and \ref{lm:6}. By Lemma \ref{lm:7}, we prove that OMLA is $\Delta/(3\Delta-1)$-competitive for the limited rejection case, where $\Delta=\max_{u\in U}\Delta_u$ (Theorem \ref{tr:2}).

\subsection{Reference System}\label{sec:4.1}

It is not straightforward to directly analyze the performance of the original system, so that we need to find an intermediate value to bound it. One key challenge of the original problem is introduced by different processing levels.  
With controllable processing time, $L$ processing levels form one additional dimension. The reference system is to construct another bipartite matching system without this additional dimension, to provide a lower bound of the original tripartite matching system. We will also need to show: 1) the reference system is a valid system; and 2) the expected reward of this reference system is a valid lower bound for $R^{\Delta_u}_{u,1}$.

For each machine $u$, we construct a reference system for $u$ as follows. The reference system has a bipartite graph $G'_u=(U'_u,V'_u;E'_u,)$, where $U'_u$ contains only one machine $u$, $V'_u$ is a set of non-repeatable tasks. At each $t$, one of $L$ tasks may come, denoted by $v'_{u,l,t}$, with a probability $p'_{u,l,t}$. Each of $L\times T$ tasks in $V'_u$ is different from each other. $v'_{u,l,t}$ can only be processed with processing level $l$. Each $v'_{u,l,t}$ has an edge to $u$. If task $v'_{u,l,t}$ arrives at $t$ and $u$ is available, we must choose $u$ and we must decide whether to assign $u$ to $v'_{u,l,t}$. Suppose $u$'s remaining budget is $\delta$ at $t$. Similar to the baseline value and activation value (Section \ref{sec:3.4}), we define the reference baseline value (resp. the reference activation value) as $\Tilde{R}^\delta_{u,t}$ (resp., $\Tilde{Q}^\delta_{u,l,t}$). The calculation of reference baseline values and reference activation values is similar to that of baseline values and activation values,
and will be given shortly. We compare the reference activation value of $u$ at $t$ and the reference baseline value of $u$ at $t+1$. The higher one indicates our choice: If the former one is larger, we assign $u$ to $v'_{u,l,t}$; Otherwise, we discard $v'_{u,l,t}$. The probability that $u$ accepts this assignment is $q'_{u,l,t}$. The reward of processing $v'_{u,l,t}$ is $r'_{u,l,t}$. The processing time $d'_l$ is drawn from the known distribution $D_l$. The rejection penalty is $\theta_l$, same as in the original system. The initial budget of $u$ is $\Delta_u$, same as in the original system. We set the parameters $p'_{u,l,t}$, $q'_{u,l,t}$, and $r'_{u,l,t}$ in the reference system as follows:
The probability that $v'_{u,l,t}$ arrives at $t$: $p'_{u,l,t}=\sum_{e\in E_u}x^*_{e,l,t}$;
The probability that $u$ accepts the assignment of task $v'_{u,l,t}$: $q'_{u,l,t}=(\sum_{e\in E_u}q_ex^*_{e,l,t})/p'_{u,l,t}$ if $p'_{u,l,t}>0$; otherwise $q'_{u,l,t}=0$; The reward of processing $v'_{u,l,t}$ (with level $l$): $ r'_{u,l,t}=(\sum_{e\in E_u}q_er_{u,v,l}x^*_{e,l,t})/(p'_{u,l,t}q'_{u,l,t})$ if $p'_{u,l,t}q'_{u,l,t}>0$, otherwise $r'_{u,l,t}=0$.
The distribution of occupation time of processing level $l$ is $\text{Pr}\{d'_l=d\}=\text{Pr}\{d_l=d\}.$

With the above parameters, we can calculate $\Tilde{Q}^\delta_{u,l,t}$ and $\Tilde{R}^\delta_{u,t}$ by
\begin{align}
    \Tilde{R}_{u,t}^\delta=&\sum_{l\in\mathcal{L}}p'_{u,l,t}\max\{\Tilde{Q}^\delta_{u,l,t},\Tilde{R}^\delta_{u,t+1}\}
    +\Big(1-\sum_{l\in\mathcal{L}}p'_{u,l,t}\Big)\Tilde{R}^\delta_{u,t+1},\ \  (\delta>0,t\in[T])\label{eqn:18},\\
    \Tilde{Q}^\delta_{u,l,t}=&q'_{u,l,t}\Big(r'_{u,l,t}+\sum_{d'\in[T-t]}\text{Pr}\{d_l=d'\}\Tilde{R}^\delta_{u,t+d'}\Big)+(1-q'_{u,l,t})\Tilde{R}^{\delta-\theta_l}_{u,t+1},\ \ (\delta>0,t\in[T])\label{eqn:19}.
\end{align}
We set $\Tilde{R}_{u,t}^\delta=0$ and $\Tilde{Q}^\delta_{u,l,t}=0$ if $\delta\leq0$ or $t>T$. We note that we do not need to calculate the specific value of $\Tilde{Q}^\delta_{u,l,t}$ and $\Tilde{R}^\delta_{u,t}$ (no computational complexity is introduced). We only need the above values in the progress of the analysis. We note that the reference system for each $u$ is a valid system, because each $p'_{u,l,t}$ and $q'_{u,l,t}$ is a valid probability value. From (\ref{eqn:offst6}), we have $p'_{u,l,t}\leq1$ and $\sum_{l}p'_{u,l,t}\leq1$. From (\ref{eqn:offst1}), we have $q'_{u,l,t}\leq1$. Therefore, $p'_{u,l,t}$ and $q'_{u,l,t}$ are valid probability values and the reference system for each machine $u$ is a valid system.

In Lemma \ref{lm:22}, we show that for each $u$, $t$, and $\delta$, the performance of the reference system $\Tilde{R}_{u,t}^\delta$ is a lower bound of the performance of $u$ in the original system $R^\delta_{u,t}$.

\begin{lemma}
\label{lm:22}
(Reference System Bounds Original System)    $R_{u,t}^\delta\geq\Tilde{R}_{u,t}^\delta$, $\forall \delta$ and $t$.
\end{lemma}

By Lemma \ref{lm:22}, we can use the lower bound of $\Tilde{R}^\delta_{u,t}$ as a valid lower bound for $R^\delta_{u,t}$. With this reference system, we first analyze the competitive ratio when each machine has an infinite initial rejection budget, then analyze the competitive ratio when each machine has an initial rejection budget no more than $\Delta$, where $\Delta=\max_{u\in U} \Delta_u$.

\subsection{Unlimited Rejection Case}\label{sec:4.2}

In this section, we analyze the competitive ratio for the unlimited rejection case. In the unlimited rejection case, each $u$ has an infinite initial rejection budget $\Delta_u=\infty$. One straightforward way is to let $\Delta_u$ to be sufficiently large when we run Algorithms \ref{alg:online} and \ref{alg:QR}. A more efficient way is to replace $R_{u,t}^{\delta}$, $Q_{e,l,t}^\delta$, $\Tilde{R}_{u,t}^\delta$, and $\Tilde{Q}_{u,l,t}^\delta$ ($\forall \delta$) by  $R_{u,t}$, $Q_{e,l,t}$, $\Tilde{R}_{u,t}$, and $\Tilde{Q}_{u,l,t}$ respectively as they are indifferent under different $\delta$. The slightly modified Algorithms \ref{alg:online} and \ref{alg:QR} are shown in the Appendix. 




The main result of this section is that Algorithm \ref{alg:online} is $1/2$-competitive for the unlimited rejection case (Theorem \ref{tr:1}). To get this result, we first get a lower bound of the performance of the reference system (Lemma \ref{lm:3}), then construct an auxiliary inequality (Lemma \ref{lm:4}) to show that a lower bound of the competitive ratio can be obtained by the ratio between the lower bound of $\sum_{u}\Tilde{R}_{u,1}$ and the offline optimal value LP(\texttt{Off}). Finally, by Lemma \ref{lm:4}, we get the result in Theorem \ref{tr:1}.

We first show that we have a lower bound for $\Tilde{R}_{u,1}$ in Lemma \ref{lm:3}.  $\Tilde{R}_{u,1}$ is $u$'s expected sum reward at and after $t=1$ (overall expected sum reward of $u$) in the reference system.

\begin{lemma}
\label{lm:3} (Reference System Lower Bound)
    For each $u$, we have 
    \begin{equation}
        \Tilde{R}_{u,1}\geq\frac{1}{2}\sum_{t\in[T]}\sum_{l\in\mathcal{L}}\sum_{e\in E_u}q_er_{u,v,l}x^*_{e,l,t}\label{eq4:16}.
    \end{equation}
\end{lemma}

To prove Lemma \ref{lm:3}, the key step is to establish a dual LP to derive the bound.  We also utilize the property that ``the sum of maximum is no less than the maximum of sum''. Next, we show the auxiliary inequality for the unlimited case by Lemmas \ref{lm:1}--\ref{lm:3}.
\begin{lemma}
\label{lm:4} (Auxiliary  Inequality)
    For the original system, in the unlimited rejection case, we have
    \begin{align}
        \frac{\mathbb{E}_{I\sim\mathcal{I}}[\text{ALG}(I)]}{\mathbb{E}_{I\sim\mathcal{I}}[\text{OPT}(I)]}\geq\frac{\frac{1}{2}\displaystyle{\sum_{u\in U}\sum_{t\in[T]}\sum_{l\in\mathcal{L}}\sum_{e\in E_u}}q_er_{u,v,l}x^*_{e,l,t}}{\text{LP}(\texttt{Off})}.
    \end{align}
\end{lemma}

 Then we introduce Theorem \ref{tr:1}. We prove that Algorithm \ref{alg:online} is $1/2$-competitive for the unlimited rejection case.

\begin{theorem}
\label{tr:1}(Competitive Ratio of Unlimited Rejection)
    OMLA is $1/2$-competitive for the problem with unlimited rejection budget.
\end{theorem}

\subsection{Limited Rejection Case}\label{sec:4.3}

In this section, we analyze the competitive ratio for the limited rejection case. Each $u$ has a finite initial rejection budget $\Delta_u<\infty$. The main result of this subsection is that Algorithm \ref{alg:online} is $\Delta/(3\Delta-1)$-competitive for the limited rejection case, where $\Delta=\max_{u\in U}\Delta_u$ (Theorem \ref{tr:2}). To get this result, we first present performance induction inequality (Lemma \ref{lm:5}), which is used to get a lower bound of the performance of the reference system (Lemma \ref{lm:6}). Then we construct an auxiliary  inequality (Lemma \ref{lm:7}) to show that a lower bound of the competitive ratio can be obtained by the ratio between the lower bound of $\sum_u\Tilde{R}_{u,1}^{\Delta_u}$ and the offline optimal value LP(\texttt{Off}). Finally, by Lemma \ref{lm:7}, we conclude the result in Theorem \ref{tr:2}.

We first show an inequality on $\Tilde{R}_{u,t}^{\delta-\theta_l}$ and $\Tilde{R}_{u,t}^\delta$, which is used to get a lower bound of the performance of the reference system. Please note that this inequality is new for the limited rejection as we need to consider the remaining budget $\delta$ now.

\begin{lemma} \label{lm:5}(Performance Induction Inequality)
    For all $\delta$, $t$ and $l$, we have
    \begin{equation}
        \Tilde{R}_{u,t}^{\delta-\theta_l}\geq\frac{\delta-\theta_l}{\delta}\Tilde{R}_{u,t}^\delta.\label{eq3:42}
    \end{equation}
\end{lemma}

Then we show that we have a lower bound for $\Tilde{R}_{u,1}^{\Delta_u}$.
\begin{lemma} \label{lm:6}(Reference System Lower Bound)
    For each $u$, we have
    \begin{equation}
        \Tilde{R}_{u,1}^{\Delta_u}\geq\frac{\Delta_u}{3\Delta_u-1}\sum_{t\in[T]}\sum_{l\in\mathcal{L}}\sum_{e\in E_u}q_er_{u,v,l}x^*_{e,l,t}.
    \end{equation}\label{eq4:19}
\end{lemma}

 The key step is to establish a dual LP to derive the bound. By Lemma \ref{lm:6}, we can eliminate the influence of different rejection penalty values (non-homogeneous $\theta_l$) of different levels from (\ref{eq4:19}). 
In the proof, it is sufficient to derive a bound utilizing the dual LP, and the dual LP can eliminate the impact of non-homogeneous $\theta_l$ in (\ref{eqn:offst2}). 

Next, we show the auxiliary inequality for the limited rejection case by Lemmas \ref{lm:1}, \ref{lm:22}, and \ref{lm:6}. 

\begin{lemma} \label{lm:7} (Auxiliary  Inequality)
    For the original system, under the limited rejection case, we have
    \begin{equation}
        \frac{\mathbb{E}_{I\sim\mathcal{I}}[\text{ALG}(I)]}{\mathbb{E}_{I\sim\mathcal{I}}[\text{OPT}(I)]}\geq\frac{\displaystyle{\sum_{u\in U}\frac{\Delta_u}{3\Delta_u-1}\sum_{t\in[T]}\sum_{l\in\mathcal{L}}\sum_{e\in E_u}}q_er_{u,v,l}x^*_{e,l,t}}{\text{LP}(\texttt{Off})}.
    \end{equation}
\end{lemma}

Then we introduce Theorem \ref{tr:2}. We prove that OMLA is $\Delta/(3\Delta-1)$-competitive for the limited rejection case, where $\Delta=\max_{u\in U}\Delta_u$.
\begin{theorem}
\label{tr:2}
    OMLA is a $\Delta/(3\Delta-1)$-competitive algorithm for the limited rejection case, where $\Delta=\max_{u\in U}\Delta_u$.
\end{theorem}

\section{Evaluation}\label{sec:5}



\subsection{Benchmarks}\label{sec:5.1}

In this section, we evaluate OMLA against five benchmarks: 1) Random (R): the system randomly chooses a machine-level pair when an online task arrives. If the chosen machine is available, we assign this machine-level pair. Otherwise, we discard the task. 2) Utility Greedy (UG): when task $v$ arrives, the system ranks all the machine-level pairs by $r_{u,v,l}$ and chooses the highest one available. 3) Efficiency Greedy (EG): with the expectation $\mathbb{E}[d_l]$ of the occupation time of processing level $l$ calculated in advance, when task $v$ arrives, the system ranks all the machine-level pair by $r_{u,v,l}/\mathbb{E}[d_l]$ and chooses the highest one available. 4) Utility Greedy + (UG+): when a task $v$ arrives, we choose a machine $u$ by \cite{RN51}, then choose the level $l$ with the highest $r_{u,v,l}$. 5) Efficiency Greedy + (EG+): when a task $v$ arrives, we choose a machine $u$ by \cite{RN51}, then choose the level $l$ with the highest $r_{u,v,l}/\mathbb{E}[d_l]$. Please note that that \cite{RN51} did not consider processing level. We use the approach in \cite{RN51} to choose machine and then use greedy method to choose level in UG+ and EG+. 

\subsection{Synthetic Dataset}\label{sec:5.2}

We generate the synthetic data set in the experiment. (The approach was also adopted in \cite{RN51}.)  We set $|U|=10$, $|V|=25$, and $T=100$. For each $u$ and $v$, an edge $(u,v)$ exists in $E$ with probability $0.1$. For each $e\in E$, we set $q_e\sim U(0.5,1)$ and $r_{u,v,l}\sim U(a\cdot l^{0.2},a\cdot l^{0.4})$, where $a\sim U(0.5,1)$. For each $l\in\mathcal{L}$ we set the distribution $D_l$ as a binomial distribution $B(T,l^{1.2}/20)$. For settings (a), (b) in Figure 2 and (a) (b) in Figure 3, $\Delta_u$ is drawn uniformly from $[\Delta]$. We set the rejection penalty for level $l$ as $\theta_l=l+2$.

\begin{figure}[t]
    \centering
    \begin{minipage}[b]{0.33\linewidth}
    \includegraphics[width=\linewidth]{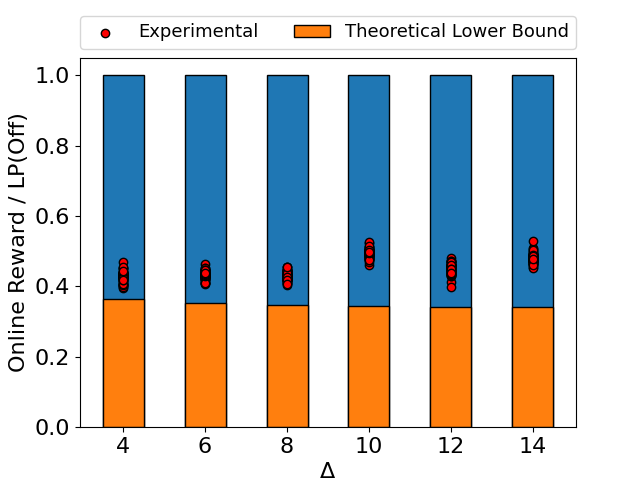}
    \subcaption{$L$=12}
    \label{fig:1b}
    \end{minipage}
    \hspace{-5.5mm}
    \begin{minipage}[b]{0.33\linewidth}
    \includegraphics[width=\linewidth]{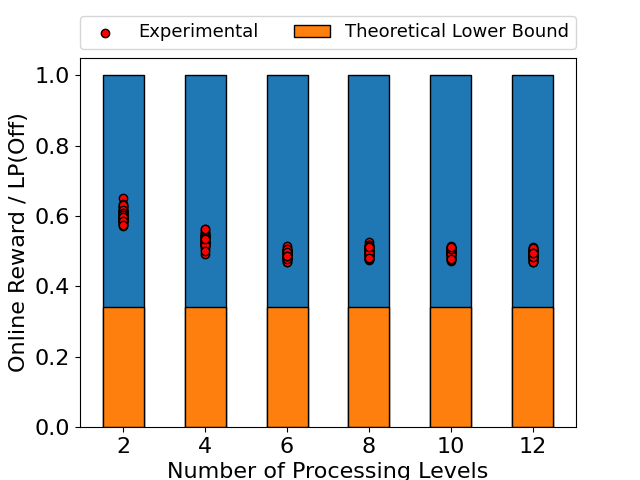}
    \subcaption{$\Delta$=20}
    \label{fig:1c}
    \end{minipage}
    \hspace{-5.5mm}
    \begin{minipage}[b]{0.33\linewidth}
    \includegraphics[width=\linewidth]{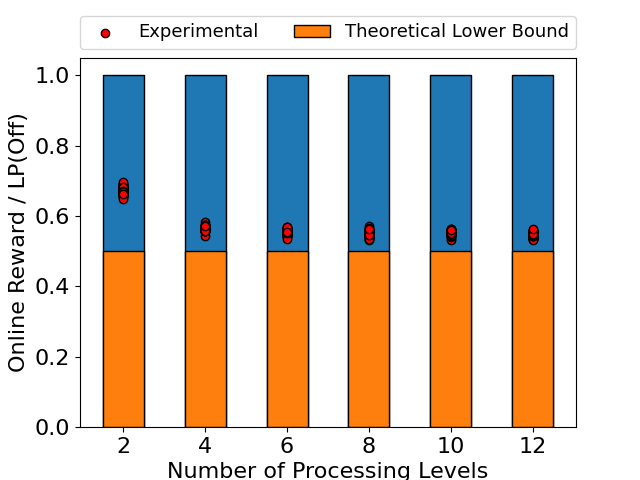}
    \subcaption{$\Delta$=$\infty$}
    \label{fig:1d}
    \end{minipage}
    
    \caption{Online Reward / LP(\texttt{Off}) of different $L$ and $\Delta$.}
    \label{fig:2}
\end{figure}

\begin{figure}[t]
    \centering
    \begin{minipage}[b]{0.33\linewidth}
    \includegraphics[width=\linewidth]{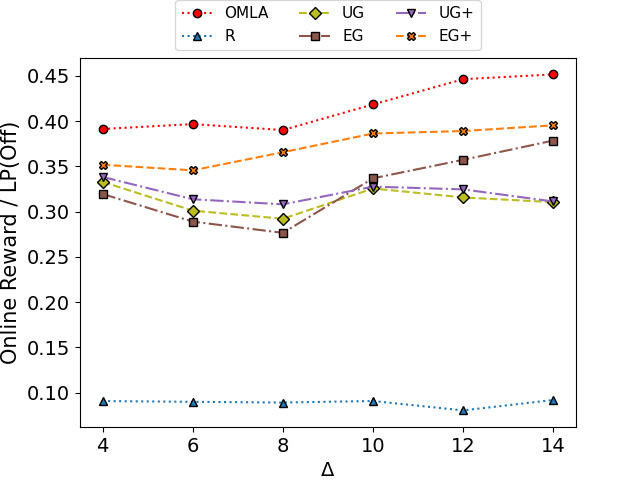}
    \subcaption{$L$=12}
    \label{fig:2b}
    \end{minipage}
    \hspace{-5.5mm}
    \begin{minipage}[b]{0.33\linewidth}
    \includegraphics[width=\linewidth]{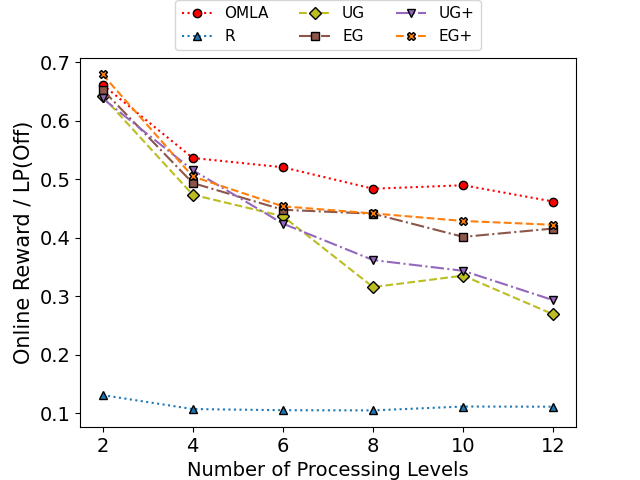}
    \subcaption{$\Delta$=20}
    \label{fig:2c}
    \end{minipage}
    \hspace{-5.5mm}
    \begin{minipage}[b]{0.33\linewidth}
    \includegraphics[width=\linewidth]{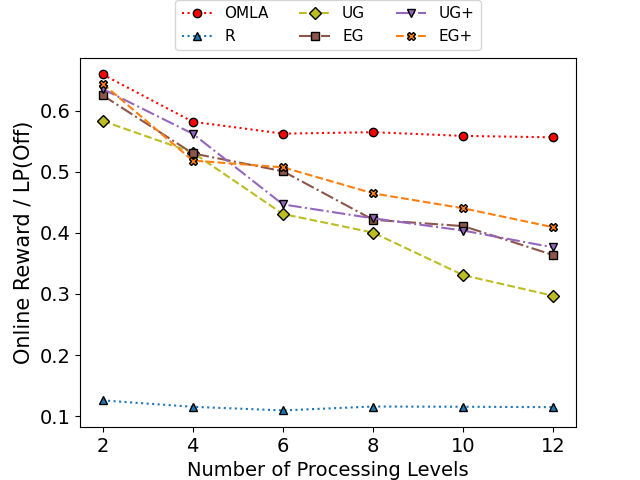}
    \subcaption{$\Delta$=$\infty$}
    \label{fig:2d}
    \end{minipage}
    
    \caption{Online Reward / LP(\texttt{Off}) of different benchmarks with different $L$ and $\Delta$.}
    \label{fig:3}
\end{figure}

\subsection{Verification of the Competitive ratio}\label{sec:5.3}

In Figure \ref{fig:2}, we investigate the ratio between the online performance of OMLA and LP(\texttt{Off}). We randomly generate a set of $\{p_{v,t}\}$, $E$ and $\{q_e\}$ for each sub-figure. For each pair of $\Delta$ and $L$, we generate a set of $\{\Delta_u\}$ and $\{r_{u,v,l}\}$, then we run $50$ rounds of experiment. In each round of experiment, we randomly generate $50$ task sequences from $\{p_{v,t}\}$, and calculate the ratio between the averaged total reward of OMLA and LP(\texttt{Off}). The orange bars in Figure \ref{fig:2} represent the competitive ratio of OMLA for each pair of $\Delta$ and $L$. The red dots in Figure \ref{fig:2} show the ratio between the averaged online total reward and LP(\texttt{Off}). Figure \ref{fig:2} shows that the ratio between the averaged performance of OMLA is indeed higher than the theoretical lower bound of the competitive ratio. The results in Figure \ref{fig:2} verifies our conclusion on the competitive ratios.

\subsection{Comparison of OMLA and Benchmarks}\label{sec:5.4}

In Figure \ref{fig:3}, we compare the performance of OMLA with benchmarks. We randomly generate a set of $\{p_{v,t}\}$, $E$ and $\{q_e\}$ for each sub-figure. For each pair of $\Delta$ and $L$, we generate a set of $\{\Delta_u\}$ and $\{r_{u,v,l}\}$. Then for each algorithm, we randomly generate $250$ task sequences from $\{p_{v,t}\}$, and calculate the ratio between the averaged total reward with LP(\texttt{Off}). OMLA outperforms all of the benchmarks with each pair of $\Delta$ and $L$. In Figures \ref{fig:2c} and \ref{fig:2d}, the performance of OMLA is slightly higher than the performance of UG+ and EG+ when the number of processing level is 2, but the performance gain becomes larger when there there are more processing levels. This demonstrates that OMLA is more advantageous for more processing levels as it is designed for joint assignment of machine and level. The results demonstrate that OMLA has the best performance with controllable processing time. OMLA provides both theoretical performance guarantees (competitive ratio) and the best average performance on the synthetic dataset.
\section{Conclusion}\label{sec:6}

In this paper, we investigate the online bipartite matching problem with controllable processing time, motivated by real-world scenarios. We design OMLA, an online algorithm to simultaneously assign an offline machine and a processing level to each online task. We prove that OMLA achieves $1/2$-competitive ratio if each machine has unlimited rejection budget and $\Delta/(3\Delta-1)$-competitive ratio if each machine has an initial budget up to $\Delta$. Furthermore, we conduct experiments on synthetic data sets, where the results demonstrate that OMLA outperforms benchmarks under a variety of environments.
\section*{Acknowledgments}

\bibliographystyle{abbrvnat}
\bibliography{main}
\clearpage
\appendix

\section{Hardness Analysis}
\begin{theorem}
No online algorithm for the problem achieves $(\frac{1}{2}+\zeta)$-competitive for any $\zeta>0$, even when $q_e=1$ for all $e\in E$ and $L=1$.
\end{theorem}
\begin{proof}

We consider the following system \cite{RN98}: $U=\{u\}$, $V=\{x,y,z\}$, $T=2$, $p_{x,1}=1$, $p_{x,2}=0$, $p_{y,1}=0$, $p_{y,2}=\epsilon$, $p_{z,1}=0$, $p_{z,2}=1-\epsilon$. $r_{u,x}=1$, $r_{u,y}=1/\epsilon$, $r_{u,z}=0$. The occupation time is equal to $T$ (Pr$\{d=T\}=1$). The acceptance probability is $q_e=1$ for all $e\in E$.

For the offline setting, there are two possible task sequences. With probability $\epsilon$, the full task sequence is $I_1=(x,y)$. With probability $1-\epsilon$, the full task sequence is $I_2=(x,z)$. For $I_1$, the offline optimal algorithm discards $x$ at $t=1$ and assigns $u$ to $y$ at $t=2$. For $I_2$, the offline optimal algorithm assigns $u$ to $x$ at $t=1$ and discards $z$ at $t=2$. Then we have $\mathbb{E}_{I\sim\mathcal{I}}[\text{OPT}(I)]=2-\epsilon$.

For any online algorithm, if it assigns $u$ to $x$ at $t=1$, it gets a total reward of $1$. If it does not assign $u$ to $x$ at $t=1$, it gets a total reward of (at most) $1/\epsilon$ with probability $\epsilon$, or it gets a total reward of $0$ with probability $1-\epsilon$. Then the expected total reward of any online algorithm is at most $1$. Thus, the competitive ratio is at most $1/(2-\epsilon)$.

\end{proof}

\section{Real-World Dataset}

\begin{figure}[b!]{}
     \centering
         \includegraphics[width=0.6\textwidth]{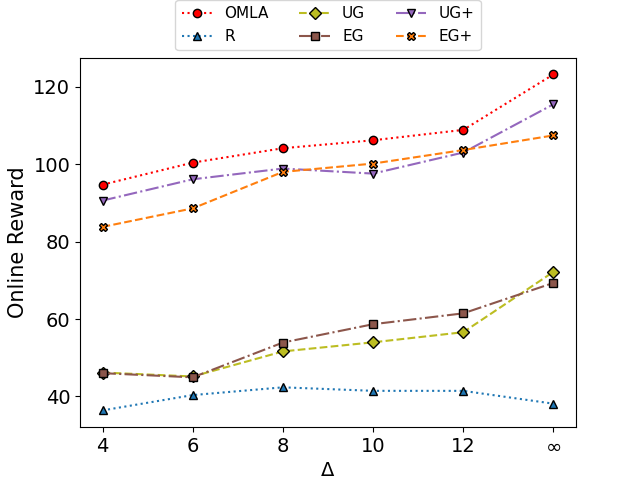}
         
         \caption{Online reward of different benchmarks with different $\Delta$, $L=3$.}
         \label{fig:4}
\end{figure}

In addition to Section \ref{sec:5.4}, we evaluate OMLA against five benchmarks using the New York City yellow cabs dataset\footnote{http://www.andresmh.com/nyctaxitrips/} (this dataset was also used by \cite{RN51} and \cite{RN50}). The dataset contains records of taxi trips in NewYork during 2013. Each trip is recorded with a taxi ID, a driver ID, the start and end location, the time and date the trip started and ended, the trip distance, the trip duration, fare amount, and toll amount. 

We extract a subset of this dataset that occurred between 7pm and 8pm on 23 weekdays in January, removing taxis with fewer than six trip records during these times. We focus on the area with longitudes from $-73\degree$ to $-75\degree$ and latitudes from $40.4\degree$ to $41\degree$, both with a step size of $0.02\degree$ (each $0.02\degree\times0.02\degree$ square is a ``grid''). In this subset of the dataset, there are 2308 trip records and 326 taxis in total. Trip records are grouped as one trip type if their start location fall within a grid and their end location also fall within a grid. After grouping, we have a total of 316 trip types. For each trip type $v$ and each taxi $u$, an edge $(u,v)$ exists if the start location of the trip type and the location of the taxi fall within a grid. We divide the one-hour period into 60 time slots, each represents a one-minute period ($T=60$). The probability $p_{v,t}$ that a trip of type $v$ arrives at $t$ is set to be proportional to the frequency that a trip of type $v$ arrives at time $t$ during 7--8PM on any one of the $23$ weekdays in January.

Of all the trips recorded in January, trips with toll amount of 0, 2.2 dollars, and 4.8 dollars account for $99.9\%$, and we choose these three toll levels as the three processing levels. For trips that can go through toll roads, we use Google Map to plan alternate routes with the above three toll levels (between 7--8PM on a weekday). Using the trip duration for different toll levels, we calculate the trip duration distribution for each level. The reward $r_{u,v,l}$ is set to the average fare of trip type $v$ minus tolls imposed by level $l$. The average fare of a trip type is set to the average fare of all trips in that trip type. We set $L=3$, where $l=1$ means taking toll roads with a fee of 4.8 dollars, $l=2$ means taking toll roads with a fee of 2.2 dollars, $l=3$ means taking no toll roads.
For the limited rejection case, the initial rejection budget $\Delta_u$ is sampled uniformly from $[\Delta]$, and the rejection penalty $\theta_l$ is set to $\theta_1=3$, $\theta_2=5$ and $\theta_3=10$. 


For each $\Delta$, we run 35 rounds of experiments. In each round of experiment, we first construct a bipartite graph $G=(U,V;E)$ as follows. We randomly sample $10$ taxis from the above $326$ taxis to form $U$, then construct $V$ as the trip types that have an edge to at least one taxi in $U$ (note that \cite{RN51} also samples a subset of taxis to reduce the size of the bipartite graph, and we improve this approach by averaging the online reward over 35 samples). For each edge $e=(u,v)$, we set the acceptance probability as $q_e\sim U(0.5,1)$. We randomly generate $40$ task sequences from the above parameters, and the average online reward of OMLA (resp. a benchmark) over these $40$ sequences is recorded as the average online reward of OMLA (resp. a benchmark) for that sample. Then, the average online reward of OMLA (resp. a benchmark) for $\Delta$ is calculated as the average reward over 35 samples.

With the above bipartite graph and the parameters, we plot the average online reward of OMLA and five benchmarks in Figure \ref{fig:4}. We notice that as $\Delta$ increases, the performance of OMLA and all the benchmarks except Random increases. This is because a larger initial rejection budget allows a taxi to reject more assignments (on average) and thus stay longer (on average) in the system, thus earning a higher average reward. As shown in Figure \ref{fig:4}, OMLA outperforms all benchmarks across all values of $\Delta$. These results demonstrate that OMLA provides the best average performance on the real-word dataset. 

\section{Complexity Analysis of Algorithm \ref{alg:online} and Algorithm \ref{alg:QR}}

In Section \ref{sec:3.4}, we proposed our OMLA in Algorithm \ref{alg:online}. At each $t$, upon the arrival of each task $v$, the complexity of Algorithm \ref{alg:online} to make a decision for this task is $\mathcal{O}(1)$. 

\begin{algorithm}[h!]
\caption{Online Machine and Level Assignment (OMLA) Algorithm for Unlimited Rejection Case}\label{alg:3}
\textbf{Input:} $U$, $V$, $E$, $\{Q_{e,l,t}\}$, $\{R_{u,t}\}$, $\mathbf{x}^*$
    \begin{algorithmic}[1]

    \For{all $t\gets1$ to $T$}
        \If{no task arrives} 
            \State skip
        \Else\ ($v$ arrives)
            \State choose pair $(u,l)$ with probability $x^*_{e,l,t}/p_{v,t}$\label{alg3:choose}
            \If{$u$ is not occupied and $Q_{e,l,t}\geq R_{u,t+1}$}\label{alg3:compare}
                \State we assign $(u,l)$ to $v$\label{alg3:9}
                \If{$u$ accepts} 
                    \State draw $d_l$ from $D_l$, $u$ gets occupied for $d_l$\label{alg3:11}
                \EndIf
            \EndIf
        \EndIf
    \EndFor
        
    \end{algorithmic}
\end{algorithm}

To execute Algorithm \ref{alg:online}, we need to run Algorithm \ref{alg:QR} in advance (in an offline manner). The complexity of Algorithm \ref{alg:QR} can be analyzed as follows. In Lines 3--5, we need at most $\Delta\cdot|U|\cdot|V|\cdot L$ steps to calculate every activation value at $t=T$. In Lines 6--16, we need at most $\Delta\cdot|U|\cdot[(L+2)\cdot |V| +2]$ steps to calculate every baseline value at $t=T$. In Lines 19--29, we have at most $|V|\cdot L\cdot (T+3)$ steps to calculate every activation values for a given set of $(\delta,u,t)$. In Line 30. we have at most $|V|\cdot L$ steps to calculate every baseline values for a given set of $(\delta,u,t)$. In Lines 17--32, we need at most $(T-1)\cdot\Delta\cdot|U|\cdot|V|\cdot L\cdot(T+4)$ steps to calculate every activation values and baseline values. As a consequence, the overall complexity of Algorithm \ref{alg:QR} is $\mathcal{O}(T^2\cdot\Delta\cdot|U|\cdot|V|\cdot L)$ (for the system with $T$ decisions to make).

\section{Modified Algorithm \ref{alg:online} and Algorithm \ref{alg:QR} for the Unlimited Rejection Case}

\begin{algorithm}[b!]
\caption{Calculation of Activation and Baseline Values for Unlimited Rejection Case}\label{alg:4}
\textbf{Input:} $U$, $V$, $E$, $\mathcal{L}$, $T$, $\{q_e\}$, $\{r_{u,v,l}\}$, $\{D_l\}$
\begin{algorithmic}[1]
\State Solve LP(\texttt{Off}) to obtain $\mathbf{x}^*$
\State $\Delta\gets\max\Delta_u$
\For{all $(u,v,l)$ that $(u,v)\in E$ and $l\in\mathcal{L}$} \label{QR4:3}
        \State $Q_{e,l,T}\gets q_er_{u,v,l}$ \label{QR4:4}
\EndFor
\For{all $u$ that $u\in U$}
    \State $a\gets0$ \label{QR4:7}
    \For{all $v$ that $(u,v)\in E_u$}
        \State $b\gets0$
        \For{all $l\in\mathcal{L}$}
            \State $b\gets b+x^*_{e,l,T}q_er_{u,v,l}$
        \EndFor
        \State $a\gets a+b$
    \EndFor
    \State $R_{u,T}\gets a$
\EndFor \label{QR4:15}
        \For{$t\gets T-1$ to $1$}
    \For{$u$ that $u\in U$}
        \For{$(v,l)$ that $(u,v)\in E_u$ and $l\in\mathcal{L}$}
            \State $a\gets0$            
                    \State $b\gets R_{u,t+1}$\label{QR4:20}
                \For{$d\gets1$ to $T-t+1$} \label{QR4:23}
                    \State $a\gets a+R_{u,t+d}$Pr$\{d_l=d\}$ \label{QR4:24}
                \EndFor\label{QR4:25}
                \State $Q_{e,l,t}\gets q_e(r_{u,v,l}+a)+(1-q_e)b$\label{QR4:26}          
    \EndFor
    \State Calculate $R_{u,t}$ by (\ref{eqn:defRift}) \label{QR4:R}
\EndFor
\EndFor
    \end{algorithmic}
    \textbf{Output:} $\{Q_{e,l,t}\}$, $\{R_{u,t}\}$, $\mathbf{x}^*$
\end{algorithm}

In this subsection, we provide the modified Algorithm \ref{alg:online} (resp. Algorithm \ref{alg:QR}) for the unlimited rejection case in Algorithm \ref{alg:3} (resp. Algorithm \ref{alg:4}). The calculation of baseline values and activation values for the unlimited rejection case is
\begin{linenomath}
\begin{align}
Q_{e,l,t}=&q_e(r_{u,v,l}+\displaystyle{\sum_{d'\in[T-t]}}\text{Pr}\{d_l=d'\}R_{u,t+d'})+(1-q_e)R_{u,t+1},\ \ (t\in[T]),\label{eqn:defQift}
\end{align}
\end{linenomath}
\begin{linenomath}
\begin{align}
    R_{u,t}=&\displaystyle{\sum_{e\in E_u}\sum_{l\in\mathcal{L}}}x^*_{e,l,t}\max\{Q_{e,l,t},R_{u,t+1}\}+(1-\displaystyle{\sum_{e\in E_u}\sum_{l\in\mathcal{L}}}x^*_{e,l,t})R_{u,t+1},(t\in[T]).\label{eqn:defRift}
\end{align}
\end{linenomath}

\section{Proof of Lemmas and Theorems}

\subsection{Proof of Lemma \ref{lm:1}}

    To achieve OPT($I$) for $I$, the offline optimal algorithm determines a probability that $u$ is assigned to $v$ with $l$ at $t$, denoted by $y_{I,e,l,t}$. The maximized expected reward OPT($I$) is equal to $\sum_{t\in[T]}\sum_{e\in E}q_e\sum_{l\in\mathcal{L}}r_{u,v,l}y_{I,e,l,t}$. The mean of $y_{I,e,l,t}$ on every sequence $I$ is $\mathbb{E}_{I\sim\mathcal{I}}[y_{I,e,l,t}]$, which is referred to as $y^*_{e,l,t}$. The offline optimal value $\mathbb{E}_{I\sim\mathcal{I}}[\text{OPT}(I)]$ is equal to $\sum_{t\in[T]}\sum_{e\in E}q_e\sum_{l\in\mathcal{L}}r_{u,v,l} y^*_{e,l,t}$. For the convenience, we define $\mathbf{y}_I\defeq\{y_{I,e,l,t}\}$ for each task sequence $I$, and $\mathbf{y}^*\defeq\{y^*_{e,l,t}\}$.
    
    To prove Lemma \ref{lm:1}, we show that each $\mathbf{y}_I$ is a valid solution to LP($\texttt{Off}$), so that $\mathbf{y}^*$ is a valid solution to LP($\texttt{Off}$). That is, they satisfy constraints (\ref{eqn:offst1}) to (\ref{eqn:offst6}).

    Given a task sequence $I$ and $\mathbf{y}_I$: \textcircled{1} for each $u$ and $t$, if $u$ is occupied by a previous task, we do not assign any new task to $u$ at $t$; if $u$ is not occupied at $t$, we can assign at most one task to $u$, so $\mathbf{y}_I$ is valid to (\ref{eqn:offst1}); \textcircled{2} for each $u$, if it gets occupied after $T$, it has at least a remaining budget of $1$ when it accepts that task; otherwise, the total rejection penalty taken by $u$ is at most $\Delta_u+\theta-1$ (gets a rejection penalty $\theta$ with a remaining budget of $1$), so $\mathbf{y}_I$ is valid to (\ref{eqn:offst2}); \textcircled{3} for each $v$ and $t$, if $v$ arrives at $t$, we assign at most one machine to $v$; if $v$ does not arrive at $t$, we do not assign any machine to $v$, so $\mathbf{y}_I$ is valid to (\ref{eqn:offst4}); \textcircled{4} for each $e=(u,v)$ and $t$, if $v$ arrives at $t$ and we assign $u$ to $v$, we assign $u$ to $v$ with at most one processing level, so $\mathbf{y}_I$ is valid to (\ref{eqn:offst5}); \textcircled{5} for each $u$ and $t$, we assign $u$ at most once at $t$, so $\mathbf{y}_I$ is valid to (\ref{eqn:offst6}). Therefore, $\mathbf{y}_I$ is a valid solution to \texttt{Off}.

    Each $y^*_{e,l,t}$ is the expected $y_{I,e,l,t}$ on $\mathcal{I}$. Because $\mathbf{y}_I$ is valid to constraints (\ref{eqn:offst1}) to (\ref{eqn:offst6}), $\mathbf{y}^*$ is also valid to constraints (\ref{eqn:offst1}) to (\ref{eqn:offst6}). With $\mathbf{y}^*$ as the solution to \texttt{Off}, the result for the objective function of \texttt{Off} is $\sum_{t\in[T]}\sum_{e\in E}q_e\sum_{l\in\mathcal{L}}r_{u,v,l} y^*_{e,l,t}$. Then the following equation holds:
    \begin{linenomath}
    \begin{align}
        \mathbb{E}_{I\sim\mathcal{I}}[\text{OPT}(I)]=&\sum_{t\in[T]}\sum_{e\in E}q_e\sum_{l\in\mathcal{L}}r_{u,v,l} y^*_{e,l,t}\nonumber\\
        \leq&\displaystyle{\sum_{t\in[T]}\sum_{e\in E}q_e\sum_{l\in\mathcal{L}}}r_{u,v,l} x^*_{e,l,t}\nonumber\\
        =&\text{LP(\texttt{Off})},\nonumber
    \end{align}
    \end{linenomath}
    where $\{x^*_{e,l,t}\}$ is the optimal solution to \texttt{Off}. Then Lemma \ref{lm:1} is proved.

\subsection{Proof of Lemma \ref{lm:22}}

    Because we set $R_{u,t}^\delta=0$ and $\Tilde{R}_{u,t}^\delta=0$ if $\delta\leq0$, $R_{u,t}^\delta\geq\Tilde{R}_{u,t}^\delta$ holds for all $t\in[T]$ when $\delta\leq0$.
    The rest of Lemma \ref{lm:22} is proved by the backward induction on $t$. We begin with the initial condition at $T$. For all $\delta>0$, we have
    \begin{equation}
        \begin{cases}
            \Tilde{Q}^\delta_{u,l,T}=q'_{u,l,T}r'_{u,l,T}=(\sum_{e\in E_u}q_er_{u,v,l}x^*_{e,l,T})/{p'_{u,l,T}},\\
            \Tilde{R}^\delta_{u,T}=\sum_{l\in\mathcal{L}}p'_{u,l,T}\Tilde{Q}^\delta_{u,l,T}=\sum_{e\in E_u}q_e\sum_{l\in \mathcal{L}}r_{u,v,l}x^*_{e,l,T}.
        \end{cases}\label{eq2:16}
    \end{equation}
    By (\ref{eqn:11}) and (\ref{eq2:16}), we have $R_{u,T}^\delta=\Tilde{R}_{u,T}^\delta$ for all $\delta$.

    For $t\in[T]$, we assume that for every $\delta$ and $t'\in[t+1,T]$, we have $R_{u,t'}^\delta\geq\Tilde{R}_{u,t'}^\delta$. By the calculation of $R^\delta_{u,t}$ (\ref{eqn:defR}), we have
    \begin{linenomath}
    \begin{align}
        R^\delta_{u,t}=&\displaystyle{\sum_{e\in E_u}\sum_{l\in\mathcal{L}}}x^*_{e,l,t}\max\{Q_{e,l,t}^\delta,R_{u,t+1}^{\delta}\}+(1-\displaystyle{\sum_{e\in E_u}\sum_{l\in\mathcal{L}}}x^*_{e,l,t})R_{u,t+1}^{\delta}\nonumber\\
        \geq&\displaystyle{\sum_{l\in\mathcal{L}}\max\{\sum_{e\in E_u}x^*_{e,l,t}Q^\delta_{e,l,t},p'_{u,l,t}R^\delta_{u,t+1}\}}
        +(1-\sum_{l\in\mathcal{L}}p'_{u,l,t})R^\delta_{u,t+1}\nonumber\\
        \geq&\displaystyle{\sum_{l\in\mathcal{L}}\max\{\sum_{e\in E_u}x^*_{e,l,t}Q^\delta_{e,l,t},p'_{u,l,t}\Tilde{R}^\delta_{u,t+1}\}}
        +(1-\sum_{l\in\mathcal{L}}p'_{u,l,t})\Tilde{R}^\delta_{u,t+1}.\label{eq2:17}
    \end{align}
    \end{linenomath}
    The last inequality in (\ref{eq2:17}) comes from the induction hypothesis, where $R_{u,t'}^\delta\geq\Tilde{R}_{u,t'}^\delta$ holds for all $\delta$ and $t'\in[t+1,T]$. For each $l\in\mathcal{L}$, we have
    \begin{linenomath}
    \begin{align}
        \sum_{e\in E_u}x^*_{e,l,t}Q^\delta_{e,l,t}=&\sum_{e\in E_u}x^*_{e,l,t}\Big[q_e\big(\sum_{d'=1}^{T-t}\text{Pr}\{d_l=d'\}R^\delta_{u,t+d'}
        +r_{u,v,l}\Big)+(1-q_e)R^{\delta-\theta_l}_{u,t+1}\big]\nonumber\\
        =&\sum_{e\in E_u}x^*_{e,l,t}q_e\sum_{d'=1}^{T-t}\text{Pr}\{d_l=d'\}R^\delta_{u,t+d'}+\sum_{e\in E_u}x^*_{e,l,t}q_er_{u,v,l}\nonumber\\
        &+\sum_{e\in E_u}x^*_{e,l,t}R^{\delta-\theta_l}_{u,t+1}
        -\sum_{e\in E_u}x^*_{e,l,t}q_eR^{\delta-\theta_l}_{u,t+1}\nonumber\\
        =&p'_{u,l,t}q'_{u,l,t}\sum_{d'=1}^{T-t}\text{Pr}\{d_l=d'\}R^\delta_{u,t+d'}
        +r'_{u,l,t}p'_{u,l,t}q'_{u,l,t}+p'_{u,l,t}R^{\delta-\theta_l}_{u,t+1}
        -p'_{u,l,t}q'_{u,l,t}R^{\delta-\theta_l}_{u,t+1}\nonumber\\
        =&p'_{u,l,t}q'_{u,l,t}\sum_{d'=1}^{T-t}\text{Pr}\{d_l=d'\}R^\delta_{u,t+d'}
        +r'_{u,l,t}p'_{u,l,t}q'_{u,l,t}
        +p'_{u,l,t}(1-q'_{u,l,t})R^{\delta-\theta_l}_{u,t+1}\nonumber\\
        \geq&p'_{u,l,t}\Big[q'_{u,l,t}\sum_{d'=1}^{T-t}\text{Pr}\{d_l=d'\}\Tilde{R}^\delta_{u,t+d'}
        +r'_{u,l,t}q'_{u,l,t}+(1-q'_{u,l,t})\Tilde{R}^{\delta-\theta_l}_{u,t+1}\Big]\nonumber\\
        =&p'_{u,l,t}\Tilde{Q}^\delta_{u,l,t}.\label{eq2:18}
    \end{align}
    \end{linenomath}
    By (\ref{eq2:17}) and (\ref{eq2:18}), we have
    \begin{linenomath}
    \begin{align}
        R^\delta_{u,t}\geq&\displaystyle{\sum_{l\in\mathcal{L}}\max\{p'_{u,l,t}\Tilde{Q}^\delta_{u,l,t},p'_{u,l,t}\Tilde{R}^\delta_{u,t+1}\}}
        +(1-\sum_{l\in\mathcal{L}}p'_{u,l,t})\Tilde{R}^\delta_{u,t+1}\nonumber\\
        =&\displaystyle{\sum_{l\in\mathcal{L}}p'_{u,l,t}\max\{\Tilde{Q}^\delta_{u,l,t},\Tilde{R}^\delta_{u,t+1}\}}
        +(1-\sum_{l\in\mathcal{L}}p'_{u,l,t})\Tilde{R}^\delta_{u,t+1}\nonumber\\
        =&\Tilde{R}^\delta_{u,t}.\label{eq2:19}
    \end{align}
    \end{linenomath}
    The last equality in (\ref{eq2:19}) comes from (\ref{eqn:18}). By (\ref{eq2:19}), induction hypothesis and the initial condition at $T$, $R_{u,t}^\delta\geq\Tilde{R}_{u,t}^\delta$ holds for all $\delta>0$ and $t\in[T]$.

    Therefore, Lemma \ref{lm:22} is proved.

\subsection{Proof of Lemma \ref{lm:3}}

To prove Lemma \ref{lm:3}, we first derive an inequality of $\Tilde{R}_{u,t}$ (\ref{eq2:21}). We construct an LP (\ref{eqn:27}) to minimize $\Tilde{R}_{u,1}$, subject to the conditions in (\ref{eq2:21}), so that we can get a lower bound of $\Tilde{R}_{u,1}$ by finding a valid solution to the dual LP (\ref{eqn:28}) of (\ref{eqn:27}). With this valid solution to the dual LP, we can have a lower bound for $\Tilde{R}_{u,1}$. We provide the detail of this proof step by step.


1) In the first step, we derive the inequality (\ref{eq2:21}) for $\Tilde{R}_{u,t}$. Similar to (\ref{eqn:18}) and (\ref{eqn:19}), the reference baseline values and reference activation values for the unlimited rejection case can be calculated as follows:
\begin{linenomath}
\begin{align}
    \Tilde{R}_{u,t}=&\sum_{l\in\mathcal{L}}p'_{u,l,t}\max\{\Tilde{Q}_{u,l,t},\Tilde{R}_{u,t+1}\}
    +(1-\sum_{l\in\mathcal{L}}p'_{u,l,t})\Tilde{R}_{u,t+1},\label{eqn:24}
\end{align}
\end{linenomath}
and
\begin{linenomath}
\begin{align}
    \Tilde{Q}_{u,l,t}=&q'_{u,l,t}(r'_{u,l,t}+\sum_{d'\in[T-t]}\text{Pr}\{d_l=d'\}\Tilde{R}_{u,t+d'})
    +(1-q'_{u,l,t})\Tilde{R}_{u,t+1},
\end{align}
\end{linenomath}
with the initial conditions
\begin{linenomath}
\begin{equation}
        \begin{cases}
            \Tilde{Q}_{u,l,T}=q'_{u,l,T}r'_{u,l,T}=(\sum_{e\in E_u}q_er_{u,v,l}x^*_{e,l,T})/{p'_{u,l,T}},\\
            \Tilde{R}_{u,T}=\sum_{l\in\mathcal{L}}p'_{u,l,T}\Tilde{Q}_{u,l,T}=\sum_{e\in E_u}q_e\sum_{l\in \mathcal{L}}r_{u,v,l}x^*_{e,l,T}.
        \end{cases}\label{eq2:16}
    \end{equation}
    \end{linenomath}
Equation (\ref{eqn:24}) implies that
\begin{linenomath}
\begin{equation}
    \Tilde{R}_{u,t}\geq\max\{\Tilde{R}_{u,t+1}+\sum_{l\in\mathcal{L}}p'_{u,l,t}(\Tilde{Q}_{u,l,t}-\Tilde{R}_{u,t+1}),\Tilde{R}_{u,t+1}\}\label{eq2:21}.
\end{equation}
\end{linenomath}
 For the convenience, we define $B_{u,l,t}\defeq p'_{u,l,t}q'_{u,l,t}$ and $A_{u,l,t}\defeq p'_{u,l,t}q'_{u,l,t}\text{Pr}\{d_l=1\}+p'_{u,l,t}(1-q'_{u,l,t})+(1/L-p'_{u,l,t})$ for each $t\in[T]$, so that we can write the first term in (\ref{eq2:21}) as 
 \begin{linenomath}
\begin{align}
    \Tilde{R}_{u,t+1}+\sum_{l\in\mathcal{L}}p'_{u,l,t}(\Tilde{Q}_{u,l,t}-\Tilde{R}_{u,t+1})
    =\sum_{l\in\mathcal{L}}B_{u,l,t}r'_{u,l,t}+\sum_{l\in\mathcal{L}}B_{u,l,t}\sum_{d'=2}^{T-t}\text{Pr}\{d_l=d'\}\Tilde{R}_{u,t+d'}+\sum_{l\in\mathcal{L}}A_{u,l,t}\Tilde{R}_{u,t+1}.
\end{align}
\end{linenomath}
Besides, $A_{u,l,t}$ can be expressed in another way
\begin{linenomath}
\begin{align}
    A_{u,l,t}&=p'_{u,l,t}q'_{u,l,t}\text{Pr}\{d_l=1\}+\frac{1}{L}-p'_{u,l,t}q'_{u,l,t}\nonumber\\
    &=\frac{1}{L}-B_{u,l,t}\text{Pr}\{d_l\geq2\}.\label{eq2:23}
\end{align}
\end{linenomath}

2) In the second step, we construct an LP (\ref{eqn:27}) to minimize $\Tilde{R}_{u,1}$ subject to the conditions in (\ref{eq2:21}) as shown below.
\begin{linenomath}
\begin{align}
    \min\ \ &\Tilde{R}_{u,1}\nonumber\\
    \text{s.t}\ \ &\Tilde{R}_{u,t}\geq\sum_{l\in\mathcal{L}}B_{u,l,t}\sum_{d'=2}^{T-t}\text{Pr}\{d_l=d'\}\Tilde{R}_{u,t+d'}+\sum_{l\in\mathcal{L}}A_{u,l,t}\Tilde{R}_{u,t+1}+\sum_{l\in\mathcal{L}}B_{u,l,t}r'_{u,l,t},\ \ (t\in[T-1]),\nonumber\\
    &\Tilde{R}_{u,t}\geq \Tilde{R}_{u,t+1},\ \ (t\in[T-1]),\nonumber\\
    &\Tilde{R}_{u,T}\geq\sum_{l\in\mathcal{L}}B_{u,l,T}r'_{u,l,T},\nonumber\\
    &\Tilde{R}_{u,t}\geq0,\ \ (t\in[T]).
    \label{eqn:27}
\end{align}
\end{linenomath}
Note that we do not need to calculate the specific solution to (\ref{eqn:27}). We only need the above LP for analysis. The minimization problem in (\ref{eqn:27}) is a LP problem, and its dual LP is given by
\begin{linenomath}
\begin{align}
    \max\ \ &\sum_{t\in[T]}\alpha_{u,t}\sum_{l\in\mathcal{L}}B_{u,l,t}r'_{u,l,t}\nonumber\\
    \text{s.t.}\ \ &\alpha_{u,1}+\beta_{u,1}\leq1,\nonumber\\
    &\alpha_{u,2}+\beta_{u,2}\leq\alpha_{u,1}\sum_{l\in\mathcal{L}}A_{u,l,1}+\beta_{u,1},\nonumber\\
    &\alpha_{u,t}+\beta_{u,t}\leq\sum_{t'\in[t-2]}\alpha_{u,t'}\sum_{l\in\mathcal{L}}B_{u,l,t'}\text{Pr}\{d_l=t-t'\}+\alpha_{u,t-1}\sum_{l\in\mathcal{L}}A_{u,l,t-1}+\beta_{u,t-1},\ \ (t\in[3,T-1]),\nonumber\\
    &\alpha_{u,T}\leq\sum_{t'\in[T-2]}\alpha_{u,t'}\sum_{l\in\mathcal{L}}B_{u,l,t'}\text{Pr}\{d_l=T-t'\}+\beta_{u,T-1}+\alpha_{u,T-1}\sum_{l\in\mathcal{L}}A_{u,l,T-1},\nonumber\\
    &\alpha_{u,t}\geq0,\ \ (t\in[T]),\nonumber\\
    &\beta_{u,t}\geq0,\ \ (t\in[T-1]).\label{eqn:28}
\end{align}
\end{linenomath}

3) In the third step, to get a lower bound on (\ref{eqn:27}), we construct a valid solution to its dual LP (\ref{eqn:28}). We set
\begin{linenomath}
\begin{equation}
    \begin{cases}
        \alpha_{u,t}=\gamma,\quad\quad\quad\quad\quad\quad\quad\quad\quad\quad\quad\quad\quad\quad\quad\quad t\in[T],\\
        \beta_{u,1}=1-\gamma,\\
        \beta_{u,2}=\beta_{u,1}-\gamma+\gamma\sum_{l\in\mathcal{L}}A_{u,l,1},\\
        \beta_{u,t}=\gamma\sum_{l\in\mathcal{L}}\sum_{t'\in[t-2]}B_{u,l,t'}\text{Pr}\{d_l=t-t'\}\\
        \quad\quad+\beta_{u,t-1}-\gamma+\gamma\sum_{l\in\mathcal{L}}A_{u,l,t-1},\quad\quad t\in[3,T-1],
    \end{cases}\label{eqn:29}
\end{equation}
\end{linenomath}
and we show that for $t\geq2$, $\beta_{u,t}$ can be calculated by
\begin{linenomath}
\begin{equation}
    \beta_{u,t}=1-\gamma-\gamma\sum_{l\in\mathcal{L}}\sum_{t'<t}B_{u,l,t'}\text{Pr}\{d_l\geq t-t'+1\}\label{eq3:27}.
\end{equation}
\end{linenomath}

            First, when $t=2$, we have
            \begin{linenomath}
            \begin{align}
                \beta_{u,2}=&1-\gamma-\gamma+\gamma\sum_{l\in\mathcal{L}}A_{u,l,1}\nonumber\\
                =&1-\gamma-\gamma+(1-\sum_{l\in\mathcal{L}}B_{u,l,1}\text{Pr}\{d_l\geq2\})\gamma\nonumber\\
                =&1-\gamma-\gamma\sum_{l\in\mathcal{L}}\sum_{t'<2}B_{u,l,t'}\text{Pr}\{d_l\geq 2-t'+1\}.\label{eq3:28}
            \end{align}
            \end{linenomath}
            The second equation of (\ref{eq3:28}) comes from (\ref{eq2:23}). This shows that (\ref{eq3:27}) is valid when $t=2$. 
            
            Next, We prove the rest of (\ref{eq3:27}) by induction. For $t\geq3$, we assume that
            \begin{linenomath}
            \begin{equation}
                \beta_{u,t-1}=1-\gamma-\gamma\sum_{l\in\mathcal{L}}\sum_{t'<t-1}B_{u,l,t'}\text{Pr}\{d_l\geq t-t'\}.
            \end{equation}
            \end{linenomath}
            Then by (\ref{eqn:29}), for $\beta_{u,t}$ we have
            \begin{linenomath}
            \begin{align}
                \beta_{u,t}=&1-\gamma-\gamma\sum_{l\in\mathcal{L}}\sum_{t'<t-1}B_{u,l,t'}\text{Pr}\{d_l\geq t-t'\}+\gamma\sum_{l\in\mathcal{L}}\sum_{t'\in[t-2]}B_{u,l,t'}\text{Pr}\{d_l=t-t'\}-\Big(1-\sum_{l\in\mathcal{L}}A_{u,l,t-1}\Big)\gamma.\label{eq3:30}
            \end{align}
            \end{linenomath}
            By (\ref{eq2:23}) and (\ref{eq3:30}), we have
            \begin{linenomath}
            \begin{align}
                \beta_{u,t}=&1-\gamma-\gamma\sum_{l\in\mathcal{L}}B_{u,l,t-1}\text{Pr}\{d_l\geq2\}-\gamma\sum_{l\in\mathcal{L}}\sum_{t'<t-1}B_{u,l,t'}\text{Pr}\{d_l\geq t-t'\}\nonumber\\
                &+\gamma\sum_{l\in\mathcal{L}}\sum_{t'<t-1}B_{u,l,t'}\text{Pr}\{d_l=t-t'\}\nonumber\\
                =&1-\gamma-\gamma\sum_{l\in\mathcal{L}}B_{u,l,t-1}\text{Pr}\{d_l\geq2\}-\gamma\sum_{t'<t-1}\sum_{l\in\mathcal{L}}B_{u,l,t'}\text{Pr}\{d_l\geq t-t'+1\}\nonumber\\
                =&1-\gamma-\gamma\sum_{t'<t}\sum_{l\in\mathcal{L}}B_{u,l,t'}\text{Pr}\{d_l\geq t-t'+1\}.\label{eq3:31}
            \end{align}
            \end{linenomath}
            Then by (\ref{eq3:31}), the induction hypothesis and the initial condition at $\beta_{u,2}$, (\ref{eq3:27}) is valid when $t\geq2$.
        
4) In the fourth step, we show that when $\gamma\in[0,1/2]$, $\alpha_{u,t}$ and $\beta_{u,t}$ defined in (\ref{eqn:29}) is a valid solution to (\ref{eqn:28}).

        First, with $\gamma\in[0,1/2]$ and (\ref{eqn:29}), it is straightforward that $\alpha_{u,t}$ and $\beta_{u,t}$ are valid to the first, second, third and fifth constraints in the dual LP (\ref{eqn:28}). 
        
        Next, we show that $\beta_{u,t}\geq0$ for all $t$ (the last constraint in (\ref{eqn:28})). For $t\in[2,T-1]$, by (\ref{eq3:27}) and $B_{u,l,t}=\sum_{e\in E_u}q_ex^*_{e,l,t}$, we have
        \begin{linenomath}
        \begin{equation}
            \beta_{u,t}=1-\gamma-\gamma\sum_{t'<t}\sum_{l\in\mathcal{L}}\text{Pr}\{d_l\geq t-t'+1\}\sum_{e\in E_u}q_ex^*_{e,l,t'}.\label{eq3:32}
        \end{equation}
        \end{linenomath}
        By (\ref{eqn:offst1}) and (\ref{eq3:32}), we have $\beta_{u,t}\geq1-2\gamma$.
        Therefore, with $\gamma\in[0,1/2]$, we have $\beta_{u,t}\geq1-2\gamma\geq0$ and $\beta_{u,t}$ defined in (\ref{eqn:29}) satisfy the last constraint in (\ref{eqn:28}).

        Then, we show that the fourth constraint in (\ref{eqn:28}) is satisfied. With $\alpha_{u,t}=\gamma$, the fourth constraint in (\ref{eqn:28}) is equivalent to
        \begin{linenomath}
        \begin{align}
            \gamma\ \leq\ &\gamma\sum_{t'\in[T-2]}\sum_{l\in\mathcal{L}}B_{u,l,t'}\text{Pr}\{d_l=T-t'\}+\beta_{u,T-1}+\gamma\sum_{l\in\mathcal{L}}A_{u,l,T-1}.\label{eq3:34}
        \end{align}
        \end{linenomath}
        By (\ref{eq2:23}) and (\ref{eq3:27}), we can reorganize the right hand side (RHS) of (\ref{eq3:34}) as
        \begin{linenomath}
        \begin{align}
            \text{RHS}=&1-\gamma-\gamma\sum_{l\in\mathcal{L}}\sum_{t'<T-1}B_{u,l,t'}\text{Pr}\{d_l\geq T-t'+1\}+\gamma-\gamma\sum_{l\in\mathcal{L}}B_{u,l,T-1}\text{Pr}\{d_l\geq T-(T-1)+1\}\nonumber\\
            =&1-\gamma\sum_{l\in\mathcal{L}}\sum_{t'<T}B_{u,l,t'}\text{Pr}\{d_l\geq T-t'+1\}.\label{eq3:35}
        \end{align}
        \end{linenomath}
        By (\ref{eq3:35}), (\ref{eq3:34}) is equivalent to
        \begin{linenomath}
        \begin{equation}
            1\geq\gamma(1+\sum_{l\in\mathcal{L}}\sum_{t'<T}B_{u,l,t'}\text{Pr}\{d_l\geq T-t'+1\}).\label{eq3:36}
        \end{equation}
        \end{linenomath}
        With $B_{u,l,t}=\sum_{e\in E_u}q_ex^*_{e,l,t}$, (\ref{eq3:34}) is satisfied if
        \begin{linenomath}
        \begin{equation}
            1\geq\gamma(1+\sum_{l\in\mathcal{L}}\sum_{t'<T}\sum_{e\in E_u}q_ex^*_{e,l,t'}\text{Pr}\{d_l\geq T-t'+1\}).\label{eq3:37}
        \end{equation}
        \end{linenomath}
        By (\ref{eqn:offst1}) we have the following inequality for the RHS of (\ref{eq3:37})
        \begin{linenomath}
        \begin{equation}
            \text{RHS of (\ref{eq3:37})}\leq2\gamma,
        \end{equation}
        \end{linenomath}
        so (\ref{eq3:34}) is satisfied if $1\geq2\gamma$. When $\gamma\in[0,1/2]$, $1\geq2\gamma$ is valid. $\alpha_{u,t}$ and $\beta_{u,t}$ defined in (\ref{eqn:29}) satisfy the fourth constraint in (\ref{eqn:28}).

        We have demonstrated that when $\gamma\in[0,1/2]$, $\alpha_{u,t}$ and $\beta_{u,t}$ defined in (\ref{eqn:29}) is a valid solution to (\ref{eqn:28}), because all the constraints in (\ref{eqn:28}) are satisfied.

5) In the fifth step, we provide a valid lower bound for $\Tilde{R}_{u,1}$ with $\alpha_{u,t}$ and $\beta_{u,t}$ defined in (\ref{eqn:29}). With $\gamma=1/2$, $\alpha_{u,t}$ and $\beta_{u,t}$ defined in (\ref{eqn:29}) is a valid solution to (\ref{eqn:28}), and the result for the object function of (\ref{eqn:28}) is
\begin{linenomath}
\begin{align}
    \frac{1}{2}\sum_{t\in[T]}\sum_{l\in\mathcal{L}}B_{u,l,t}r'_{u,l,t}=&\frac{1}{2}\sum_{t\in[T]}\sum_{l\in\mathcal{L}}\sum_{e\in E_u}q_er_{u,v,l}x^*_{e,l,t}\nonumber\\
    \leq&\max\sum_{t\in[T]}\alpha_{u,t}\sum_{l\in\mathcal{L}}B_{u,l,t}r'_{u,l,t}.\label{eqn:30}
\end{align}
\end{linenomath}

According to the linear programming duality theorem, we have
\begin{linenomath}
\begin{equation}
    \min \Tilde{R}_{u,1}=\max\sum_{t\in[T]}\alpha_{u,t}\sum_{l\in\mathcal{L}}B_{u,l,t}r'_{u,l,t},\label{eq2:26}
\end{equation}
\end{linenomath}
and
\begin{linenomath}
\begin{equation}
    \min \Tilde{R}_{u,1}\geq\frac{1}{2}\sum_{t\in[T]}\sum_{l\in\mathcal{L}}\sum_{e\in E_u}q_er_{u,v,l}x^*_{e,l,t}.
\end{equation}
\end{linenomath}

Therefore, Lemma \ref{lm:3} is proved.

\subsection{Proof of Lemma \ref{lm:4}}

By Lemma \ref{lm:1}, we have
\begin{linenomath}
\begin{equation}
    \text{LP(\texttt{Off})}=\displaystyle{\sum_{t\in[T]}\sum_{e\in E}q_e\sum_{l\in\mathcal{L}}}r_{u,v,l} x^*_{e,l,t}\geq\mathbb{E}_{I\sim \mathcal{I}}[\text{OPT}(I)].\label{eq5:43}
\end{equation}
\end{linenomath}
By Lemmas \ref{lm:22} and \ref{lm:3}, we have
\begin{linenomath}
\begin{equation}
    R_{u,1}\geq\Tilde{R}_{u,1}\geq\frac{1}{2}\sum_{t\in[T]}\sum_{l\in\mathcal{L}}\sum_{e\in E_u}q_er_{u,v,l}x^*_{e,l,t}.\label{eq5:44}
\end{equation}
\end{linenomath}
By the definition of the activation value, the expected total reward of OMLA in the original system $\mathbb{E}_{I\sim\mathcal{I}}[\text{ALG}(I)]$ is equal to $\sum_{u\in U}R_{u,1}$ (for the unlimited rejection case). Then by (\ref{eq5:43}) and (\ref{eq5:44}), we have
\begin{linenomath}
\begin{align}
    \frac{\mathbb{E}_{I\sim\mathcal{I}}[\text{ALG}(I)]}{\mathbb{E}_{I\sim\mathcal{I}}[\text{OPT}(I)]}\geq&\frac{\displaystyle{\sum_{u\in U}}R_{u,1}}{\text{LP(\texttt{Off})}}\nonumber\\
    \geq&\frac{\frac{1}{2}\displaystyle{\sum_{u\in U}\sum_{t\in[T]}\sum_{l\in\mathcal{L}}\sum_{e\in E_u}}q_er_{u,v,l}x^*_{e,l,t}}{\text{LP(\texttt{Off})}}.\label{eq2:28}
\end{align}
\end{linenomath}
Lemma \ref{lm:4} is proved.

\subsection{Proof of Theorem \ref{tr:1}}

    By Lemma \ref{lm:4}, we have
    \begin{linenomath}
    \begin{align}
        \frac{\mathbb{E}_{I\sim\mathcal{I}}[\text{ALG}(I)]}{\mathbb{E}_{I\sim\mathcal{I}}[\text{OPT}(I)]}\geq&\frac{\frac{1}{2}\displaystyle{\sum_{u\in U}\sum_{t\in[T]}\sum_{l\in\mathcal{L}}\sum_{e\in E_u}}q_er_{u,v,l}x^*_{e,l,t}}{\text{LP}(\texttt{Off})}\nonumber\\
        =&\frac{\frac{1}{2}\displaystyle{\sum_{u\in U}\sum_{t\in[T]}\sum_{l\in\mathcal{L}}\sum_{e\in E_u}}q_er_{u,v,l}x^*_{e,l,t}}{\displaystyle{\sum_{u\in U}\sum_{t\in[T]}\sum_{l\in\mathcal{L}}\sum_{e\in E_u}}q_er_{u,v,l}x^*_{e,l,t}}\nonumber\\
        =&\frac{1}{2}.
    \end{align}
    \end{linenomath}
    Theorem \ref{tr:1} is proved.

\subsection{Proof of Lemma \ref{lm:5}}

    When $\delta\leq0$, we have $\Tilde{R}_{u,t}^{\delta-\theta_l}=0$ and $\Tilde{R}_{u,t}^\delta=0$, so (\ref{eq3:42}) is valid for all $t\in[T]$ and $l\in\mathcal{L}$.  When $\delta>0$ and $\delta\leq\theta_l$, we have $\Tilde{R}_{u,t}^{\delta-\theta_l}=0$ and $\frac{\delta-\theta_l}{\delta}\Tilde{R}_{u,t}^\delta\leq0$, so (\ref{eq3:42}) is valid for all $t\in[T]$.

    Next, we prove Lemma \ref{lm:5} for $\delta>\theta_l$ by induction. The initial condition is at $t=T$. We have $\Tilde{R}^\delta_{u,T}=\Tilde{R}^{\delta'}_{u,T}=\sum_{l\in\mathcal{L}}p'_{u,l,T}q'_{u,l,T}r'_{u,l,T}$ for all $\delta,\delta'>0$, so (\ref{eq3:42}) holds for all $t=T$, $\delta>\theta_l$.

    For $t<T$, we assume that (\ref{eq3:42}) is valid for all $t'>t$ and $\delta>\theta_l$. Then by (\ref{eqn:18}), we have
    \begin{linenomath}
    \begin{align}
        \Tilde{R}_{u,t}^\delta=&\sum_{l\in\mathcal{L}}p'_{u,l,t}\max\{\Tilde{Q}^\delta_{u,l,t},\Tilde{R}^\delta_{u,t+1}\}+\Big(1-\sum_{l\in\mathcal{L}}p'_{u,l,t}\Big)\Tilde{R}^\delta_{u,t+1}\nonumber\\
        =&\sum_{l\in\mathcal{L}}\max\{p'_{u,l,t}\Tilde{Q}^\delta_{u,l,t}+(\frac{1}{L}-p'_{u,l,t})\Tilde{R}^\delta_{u,t+1},\frac{1}{L}\Tilde{R}^\delta_{u,t+1}\}\nonumber\\
        \geq&\sum_{l\in\mathcal{L}}\max\{\frac{\delta}{\delta+\theta_l}[p'_{u,l,t}\Tilde{Q}^\delta_{u,l,t}+(\frac{1}{L}-p'_{u,l,t})\Tilde{R}^\delta_{u,t+1}]+\frac{\theta_l}{\delta+\theta_l}\cdot\frac{1}{L}\Tilde{R}^\delta_{u,t+1},\frac{1}{L}\Tilde{R}^\delta_{u,t+1}\}.\label{eq3:43}
    \end{align}
    \end{linenomath}
    By (\ref{eqn:19}) and the induction hypothesis, we have
    \begin{linenomath}
    \begin{align}
        \Tilde{Q}^\delta_{u,l,t}\geq&q'_{u,l,t}\Big(r'_{u,l,t}+\sum_{d'\in[T-t]}\text{Pr}\{d_l=d'\}\Tilde{R}^\delta_{u,t+d'}\Big)+(1-q'_{u,l,t})\frac{\delta-\theta_l}{\delta}\Tilde{R}_{u,t+1}^\delta,
    \end{align}
    \end{linenomath}
    then in (\ref{eq3:43}) we have
    \begin{linenomath}
    \begin{align}
        &\frac{\delta}{\delta+\theta_l}[p'_{u,l,t}\Tilde{Q}^\delta_{u,l,t}+(\frac{1}{L}-p'_{u,l,t})\Tilde{R}^\delta_{u,t+1}]+\frac{\theta_l}{\delta+\theta_l}\cdot\frac{1}{L}\Tilde{R}^\delta_{u,t+1}\nonumber\\
        \geq&\frac{\delta}{\delta+\theta_l}\Bigg[p'_{u,l,t}\Bigg(q'_{u,l,t}\Big(r'_{u,l,t}+\sum_{d'=1}^{T-t}\text{Pr}\{d_l=d'\}\Tilde{R}^\delta_{u,t+d'}\Big)+(1-q'_{u,l,t})\frac{\delta-\theta_l}{\delta}\Tilde{R}_{u,t+1}^\delta\Bigg)+(\frac{1}{L}-p'_{u,l,t})\Tilde{R}^\delta_{u,t+1}\Bigg]\nonumber\\
        &+\frac{\theta_l}{\delta+\theta_l}\cdot\frac{1}{L}\Tilde{R}^\delta_{u,t+1}\nonumber\\
        =&\frac{\delta}{\delta+\theta_l}p'_{u,l,t}q'_{u,l,t}\sum_{d'=1}^{T-t}\text{Pr}\{d_l=d'\}\Tilde{R}^\delta_{u,t+d'}+\frac{\delta}{\delta+\theta_l}p'_{u,l,t}(1-q'_{u,l,t})\frac{\delta-\theta_l}{\delta}\Tilde{R}_{u,t+1}^\delta+\frac{\delta}{\delta+\theta_l}(\frac{1}{L}-p'_{u,l,t})\Tilde{R}^\delta_{u,t+1}\nonumber\\
        &+\frac{\delta}{\delta+\theta_l}p'_{u,l,t}q'_{u,l,t}r'_{u,l,t}+\frac{\theta_l}{\delta+\theta_l}p'_{u,l,t}q'_{u,l,t}\sum_{t'\geq1}\text{Pr}\{d_l=t'\}\Tilde{R}^\delta_{u,t+1}+\frac{\theta_l}{\delta+\theta_l}p'_{u,l,t}(1-q'_{u,l,t})\Tilde{R}^\delta_{u,t+1}\nonumber\\
        &+\frac{\theta_l}{\delta+\theta_l}(\frac{1}{L}-p'_{u,l,t})\Tilde{R}^\delta_{u,t+1}\nonumber\\
        =&\frac{\delta}{\delta+\theta_l}p'_{u,l,t}q'_{u,l,t}r'_{u,l,t}+(\frac{1}{L}-p'_{u,l,t})\Tilde{R}^\delta_{u,t+1}+p'_{u,l,t}q'_{u,l,t}\sum_{t'\geq1}\text{Pr}\{d_l=t'\}\Big(\frac{\theta_l}{\delta+\theta_l}\Tilde{R}_{u,t+1}^\delta+\frac{\delta}{\delta+\theta_l}\Tilde{R}_{u,t+t'}^\delta\Big)\nonumber\\
        &+p'_{u,l,t}(1-q'_{u,l,t})\frac{\delta}{\delta+\theta_l}\Tilde{R}^\delta_{u,t+1}.\label{eq3:45}
    \end{align}
    \end{linenomath}
    By (\ref{eqn:18}), we have $\Tilde{R}_{u,t+1}^\delta\geq\Tilde{R}_{u,t+t'}^\delta$ when $t'\geq1$. So we continue from the last equation in (\ref{eq3:45})
    \begin{linenomath}
    \begin{align}
        &\frac{\delta}{\delta+\theta_l}[p'_{u,l,t}\Tilde{Q}^\delta_{u,l,t}+(\frac{1}{L}-p'_{u,l,t})\Tilde{R}^\delta_{u,t+1}]+\frac{\theta_l}{\delta+\theta_l}\cdot\frac{1}{L}\Tilde{R}^\delta_{u,t+1}\nonumber\\
        \geq&\frac{\delta}{\delta+\theta_l}p'_{u,l,t}q'_{u,l,t}r'_{u,l,t}+(\frac{1}{L}-p'_{u,l,t})\Tilde{R}^\delta_{u,t+1}+p'_{u,l,t}q'_{u,l,t}\sum_{t'\geq1}\text{Pr}\{d_l=t'\}\Big(\frac{\theta_l}{\delta+\theta_l}\Tilde{R}_{u,t+1}^\delta+\frac{\delta}{\delta+\theta_l}\Tilde{R}_{u,t+t'}^\delta\Big)\nonumber\\
        &+p'_{u,l,t}(1-q'_{u,l,t})\frac{\delta}{\delta+\theta_l}\Tilde{R}^\delta_{u,t+1}\nonumber\\
        \geq&\frac{\delta}{\delta+\theta_l}p'_{u,l,t}q'_{u,l,t}r'_{u,l,t}+(\frac{1}{L}-p'_{u,l,t})\Tilde{R}^\delta_{u,t+1}+p'_{u,l,t}q'_{u,l,t}\sum_{t'\geq1}\text{Pr}\{d_l=t'\}\Tilde{R}_{u,t+t'}^\delta\nonumber\\
        &+p'_{u,l,t}(1-q'_{u,l,t})\frac{\delta}{\delta+\theta_l}\Tilde{R}^\delta_{u,t+1}\nonumber\\
        \geq&\frac{\delta}{\delta+\theta_l}\Big[p'_{u,l,t}q'_{u,l,t}r'_{u,l,t}+(\frac{1}{L}-p'_{u,l,t})\Tilde{R}^{\delta+\theta_l}_{u,t+1}+p'_{u,l,t}q'_{u,l,t}\sum_{t'\geq1}\text{Pr}\{d_l=t'\}\Tilde{R}_{u,t+t'}^{\delta+\theta_l}+p'_{u,l,t}(1-q'_{u,l,t})\Tilde{R}^\delta_{u,t+1}\Big]\nonumber\\
        =&\frac{\delta}{\delta+\theta_l}\Big[p'_{u,l,t}\Big(q'_{u,l,t}\big(r'_{u,l,t}+\sum_{t'\geq1}\text{Pr}\{d_l=t'\}\Tilde{R}_{u,t+t'}^{\delta+\theta_l}\big)+(1-q'_{u,l,t})\Tilde{R}^\delta_{u,t+1}\Big)+(\frac{1}{L}-p'_{u,l,t})\Tilde{R}^{\delta+\theta_l}_{u,t+1}\Big]\nonumber\\
        =&\frac{\delta}{\delta+\theta_l}\Big[p'_{u,l,t}\Tilde{Q}^{\delta+\theta_l}_{u,l,t}+(\frac{1}{L}-p'_{u,l,t})\Tilde{R}^{\delta+\theta_l}_{u,t+1}\Big].\label{eq3:46}
    \end{align}
    \end{linenomath}
    The last equation of (\ref{eq3:46}) comes from (\ref{eqn:19}). By (\ref{eq3:46}) and (\ref{eq3:43}), we have
    \begin{linenomath}
    \begin{align}
        \Tilde{R}^\delta_{u,t}\geq&\sum_{l\in\mathcal{L}}\max\Big\{\frac{\delta}{\delta+\theta_l}\Big[p'_{u,l,t}\Tilde{Q}^{\delta+\theta_l}_{u,l,t}+(\frac{1}{L}-p'_{u,l,t})\Tilde{R}^{\delta+\theta_l}_{u,t+1}\Big],\frac{1}{L}\Tilde{R}_{u,t+1}^\delta\Big\}\nonumber\\
        \geq&\frac{\delta}{\delta+\theta_l}\sum_{l\in\mathcal{L}}\max\Big\{p'_{u,l,t}\Tilde{Q}^{\delta+\theta_l}_{u,l,t}+(\frac{1}{L}-p'_{u,l,t})\Tilde{R}^{\delta+\theta_l}_{u,t+1},\frac{1}{L}\Tilde{R}_{u,t+1}^{\delta+\theta_l}\Big\}\nonumber\\
        =&\frac{\delta}{\delta+\theta_l}\Tilde{R}_{u,t}^{\delta+\theta_l}.\label{eq3:47}
    \end{align}
    \end{linenomath}

    By (\ref{eq3:47}), the induction hypothesis and the initial condition at $t=T$, we have proved that (\ref{eq3:42}) also holds for all $t\in[T]$ when $\delta>\theta_l$. Therefore, Lemma \ref{lm:5} is proved.

\subsection{Prove of Lemma \ref{lm:6}}
To prove Lemma \ref{lm:6}, we first derive an inequality of $\Tilde{R}^\delta_{u,t}$ (\ref{eqn:31}). We then construct an LP (\ref{eq2:31}) to minimize $\Tilde{R}^{\Delta_u}_{u,1}$ subject to the conditions in (\ref{eqn:31}), so that we can get a lower bound of $\Tilde{R}^{\Delta_u}_{u,1}$ by finding a valid solution to the dual LP (\ref{eq2:32}) of (\ref{eq2:31}). With this valid solution to the dual LP, we can have a lower bound for $\Tilde{R}^{\Delta_u}_{u,1}$. We provide the details of this proof step by step.

1) In the first step, we derive the inequality (\ref{eqn:31}) for $\Tilde{R}^\delta_{u,t}$. By (\ref{eqn:18}) and Lemma \ref{lm:5}, we have the following inequalities for $\Tilde{R}_{u,t}^\delta$:
\begin{linenomath}
\begin{align}
    \Tilde{R}_{u,t}^\delta\geq&\sum_{l\in\mathcal{L}}\max\Big\{p'_{u,l,t}\hat{Q}^\delta_{u,l,t}+(\frac{1}{L}-p'_{u,l,t})\Tilde{R}^\delta_{u,t+1},\frac{1}{L}\Tilde{R}^\delta_{u,t+1}\Big\}\nonumber\\
    \geq&\max\Big\{\sum_{l\in\mathcal{L}}p'_{u,l,t}\hat{Q}^\delta_{u,l,t}+(1-\sum_{l\in\mathcal{L}}p'_{u,l,t})\Tilde{R}^\delta_{u,t+1},\Tilde{R}_{u,t+1}^\delta\Big\},\label{eqn:31}
\end{align}
\end{linenomath}
where
\begin{linenomath}
\begin{align}
    \hat{Q}_{u,l,t}^\delta=&q'_{u,l,t}(r'_{u,l,t}+\sum_{d'\in[T-t']}\text{Pr}\{d_l=d'\}\Tilde{R}_{u,t+d'}^\delta)+(1-q'_{u,l,t})\frac{\delta-\theta_l}{\delta}\Tilde{R}^\delta_{u,t+1}.
\end{align}
\end{linenomath}
For convenience of notation, we define $B_{u,l,t}\defeq p'_{u,l,t}q'_{u,l,t}$ and $A_{u,l,t}\defeq p'_{u,l,t}q'_{u,l,t}$Pr$\{d_l=1\}+p'_{u,l,t}(1-q'_{u,l,t})\frac{\Delta_u-\theta_l}{\Delta_u}+(1/L-p'_{u,l,t})$ for each $t\in[T]$, so that for $\Tilde{R}^{\Delta_u}_{u,t}$ we have
\begin{linenomath}
\begin{align}
    \Tilde{R}^{\Delta_u}_{u,t}\geq&\max\Big\{\sum_{l\in\mathcal{L}}p'_{u,l,t}\hat{Q}^{\Delta_u}_{u,l,t}+(1-\sum_{l\in\mathcal{L}}p'_{u,l,t})\Tilde{R}^{\Delta_u}_{u,t+1},\Tilde{R}_{u,t+1}^{\Delta_u}\Big\}\nonumber\\
    =&\max\Big\{\sum_{l\in\mathcal{L}}B_{u,l,t}r'_{u,l,t}+\sum_{l\in\mathcal{L}}A_{u,l,t}\Tilde{R}^{\Delta_u}_{u,t+1}+\sum_{l\in\mathcal{L}}B_{u,l,t}\sum_{d'=2}^{T-t}\text{Pr}\{d_l=d'\}\Tilde{R}^{\Delta_u}_{u,t+d'},\Tilde{R}_{u,t+1}^{\Delta_u}\Big\}.
\end{align}
\end{linenomath}

2) In the second step, we construct an LP (\ref{eq2:31}) to minimize $\Tilde{R}^{\Delta_u}_{u,1}$ subject to the conditions in (\ref{eqn:31}) as shown below.
\begin{linenomath}
\begin{align}
    \min\ \ &\Tilde{R}^{\Delta_u}_{u,1}\nonumber\\
    \text{s.t}\ \ &\Tilde{R}^{\Delta_u}_{u,t}\geq\sum_{l\in\mathcal{L}}B_{u,l,t}r'_{u,l,t}+\sum_{l\in\mathcal{L}}A_{u,l,t}\Tilde{R}^{\Delta_u}_{u,t+1}+\sum_{l\in\mathcal{L}}B_{u,l,t}\sum_{d'=2}^{T-t}\text{Pr}\{d_l=d'\}\Tilde{R}^{\Delta_u}_{u,t+d'},\ (t\in[T-1]),\nonumber\\
    &\Tilde{R}^{\Delta_u}_{u,t}\geq \Tilde{R}^{\Delta_u}_{u,t+1},\ \ (t\in[T-1]),\nonumber\\
    &\Tilde{R}^{\Delta_u}_{u,T}\geq\sum_{l\in\mathcal{L}}B_{u,l,T}r'_{u,l,T},\nonumber\\
    &\Tilde{R}^{\Delta_u}_{u,t}\geq0,\ \ (t\in[T]).\nonumber\\
    \label{eq2:31}
\end{align}
\end{linenomath}
Note that we do not need to calculate the specific solution to (\ref{eq2:31}). We only need the above LP for analysis. The minimization problem in (\ref{eq2:31}) is a LP problem, and its dual LP is given by
\begin{linenomath}
\begin{align}
    \max\ \ &\sum_{t\in[T]}\alpha_{u,t}\sum_{l\in\mathcal{L}}B_{u,l,t}r'_{u,l,t}\nonumber\\
    \text{s.t.}\ \ &\alpha_{u,1}+\beta_{u,1}\leq1,\nonumber\\
    &\alpha_{u,2}+\beta_{u,2}\leq\alpha_{u,1}\sum_{l\in\mathcal{L}}A_{u,l,1}+\beta_{u,1},\nonumber\\
    &\alpha_{u,t}+\beta_{u,t}\leq\sum_{t'\in[t-2]}\alpha_{u,t'}\sum_{l\in\mathcal{L}}B_{u,l,t'}\text{Pr}\{d_l=t-t'\}+\beta_{u,t-1}+\alpha_{u,t-1}\sum_{l\in\mathcal{L}}A_{u,l,t-1},\ (t\in[3,T-1]),\nonumber\\
    &\alpha_{u,T}\leq\sum_{t'\in[T-2]}\alpha_{u,t'}\sum_{l\in\mathcal{L}}B_{u,l,t'}\text{Pr}\{d_l=T-t'\}+\beta_{u,T-1}+\alpha_{u,T-1}\sum_{l\in\mathcal{L}}A_{u,l,T-1},\nonumber\\
    &\alpha_{u,t}\geq0,\ \ (t\in[T]),\nonumber\\
    &\beta_{u,t}\geq0,\ \ (t\in[T-1]).\nonumber\\
    \label{eq2:32}
\end{align}
\end{linenomath}

3) In the third step, to get a lower bound on (\ref{eq2:31}), we construct a valid solution to its dual LP (\ref{eq2:32}). We set
\begin{linenomath}
\begin{equation}
    \begin{cases}
        \alpha_{u,t}=\gamma,\quad\quad\quad\quad\quad\quad\quad\quad\quad\quad\quad\quad\quad\quad\quad &t\in[T],\\
        \beta_{u,1}=1-\gamma,\\
        \beta_{u,2}=1-\gamma-\gamma(1-\displaystyle{\sum_{l\in\mathcal{L}}}A_{u,l,1}),\\
        \beta_{u,t}=1-\gamma-\gamma\displaystyle{\sum_{l\in\mathcal{L}}}\displaystyle{\sum_{t'<t}}p'_{u,l,t'}q'_{u,l,t'}\text{Pr}\{d_l\geq t-t'+1\}-\gamma\displaystyle{\sum_{l\in\mathcal{L}}}\frac{\theta_l}{\Delta_u}\displaystyle{\sum_{t'<t}p'_{u,l,t'}(1-q'_{u,l,t'})},\quad\quad &t\in[3,T-1],
    \end{cases}\label{eq2:33}
\end{equation}
\end{linenomath}
and we show that this is a valid solution to (\ref{eq2:32}) when $\gamma\in[0,\Delta_u/(3\Delta_u-1)]$.

    With $\gamma\in[0,\Delta_u/(3\Delta_u-1)]$ and (\ref{eq2:33}), it is straightforward that $\alpha_{u,t}$ and $\beta_{u,t}$ satisfy the first, second and fifth constraint in (\ref{eq2:32}).

    For the third constraint in (\ref{eq2:32}), we first show it is satisfied when $t=3$. When $t=3$, the third constraint in (\ref{eq2:32}) is equivalent to
    \begin{linenomath}
    \begin{align}
        \gamma+\beta_{u,3}\leq&\beta_{u,2}+\gamma\sum_{l\in\mathcal{L}}A_{u,l,2}+\gamma\sum_{t'\in[1]}\sum_{l\in\mathcal{L}}B_{u,l,t'}\text{Pr}\{d_l=3-t'\},\label{eq3:53}
    \end{align}
    \end{linenomath}
    and we have the following equations for the left hand side (LHS) and RHS of (\ref{eq3:53})
    \begin{linenomath}
    \begin{align}
        \text{LHS of (\ref{eq3:53})}=&1-\gamma\sum_{l\in\mathcal{L}}\frac{\theta_l}{\Delta_u}\sum_{t'<3}p'_{u,l,t'}(1-q'_{u,l,t'})-\gamma\sum_{l\in\mathcal{L}}\sum_{t'<3}p'_{u,l,t'}q'_{u,l,t'}\text{Pr}\{d_l\geq4-t'\},\label{eq5:59}
    \end{align}
    \end{linenomath}
    with
    \begin{linenomath}
    \begin{align}
        \text{RHS of (\ref{eq3:53})}=&1-2\gamma+\gamma\sum_{l\in\mathcal{L}}\sum_{t'<3}\Big[p'_{u,l,t'}q'_{u,l,t'}\text{Pr}\{d_l=1\}+p'_{u,l,t'}(1-q'_{u,l,t'})\frac{\Delta_u-\theta_l}{\Delta_u}+(\frac{1}{L}-p'_{u,l,t'})\Big]\nonumber\\
        &+\gamma\sum_{l\in\mathcal{L}}p'_{u,l,1}q'_{u,l,1}\text{Pr}\{d_l=2\}.\label{eq5:60}
    \end{align}
    \end{linenomath}

    By (\ref{eq5:59}) and (\ref{eq5:60}), (\ref{eq3:53}) is satisfied if
    \begin{linenomath}
    \begin{equation}
        \gamma\sum_{l\in\mathcal{L}}\sum_{t'<3}p'_{u,l,t'}q'_{u,l,t'}(\text{Pr}\{d_l\geq1\}-1)\geq0.\label{eq5:61}
    \end{equation}
    \end{linenomath}
    For all $l$, we have \text{Pr}$\{d_l\geq1\}=1$, so (\ref{eq5:61}) is satisfied, regardless of the value of $\gamma$. Then the third constraint in (\ref{eq2:32}) is satisfied when $t=3$, regardless of the value of $\gamma$.

    When $t\in[4,T-1]$, we have the following equations for the LHS and RHS of the third constraint in (\ref{eq2:32})
    \begin{linenomath}
    \begin{align}
        \text{LHS}=&1-\gamma\sum_{l\in\mathcal{L}}\frac{\theta_l}{\Delta_u}\sum_{t'<t}p'_{u,l,t'}(1-q'_{u,l,t'})-\gamma\sum_{l\in\mathcal{L}}\sum_{t'<t}p'_{u,l,t'}q'_{u,l,t'}\text{Pr}\{d_l\geq t-t'+1\},\label{eq5:62}
    \end{align}
    \end{linenomath}
    \begin{linenomath}
    \begin{align}
        \text{RHS}=&1-\gamma-\gamma\sum_{l\in\mathcal{L}}\frac{\theta_l}{\Delta_u}\sum_{t'<t-1}p'_{u,l,t'}(1-q'_{u,l,t'})-\gamma\sum_{l\in\mathcal{L}}\sum_{t'<t-1}p'_{u,l,t'}q'_{u,l,t'}\text{Pr}\{d_l\geq t-t'\}\nonumber\\
        &+\gamma\sum_{t'\in[t-2]}\sum_{l\in\mathcal{L}}p'_{u,l,t'}q'_{u,l,t'}\text{Pr}\{d_l=t-t'\}+\gamma\sum_{l\in\mathcal{L}}A_{u,l,t-1}.\label{eq5:63}
    \end{align}
    \end{linenomath}
    By (\ref{eq5:62}) and (\ref{eq5:63}), the third constraint in (\ref{eq2:32}) is satisfied if
    \begin{linenomath}
    \begin{equation}
        0\leq\gamma\sum_{l\in\mathcal{L}}p'_{u,l,t-1}q'_{u,l,t-1}(\text{Pr}\{d_l\geq1\}-1).\label{eq5:64}
    \end{equation}
    \end{linenomath}
    Because for each $l$, $\text{Pr}\{d_l\geq1\}=1$, (\ref{eq5:64}) is satisfied when $\gamma\in[0,\Delta_u/(3\Delta_u-1)]$, and thus the third constraint is satisfied when $\gamma\in[0,\Delta_u/(3\Delta_u-1)]$.

    To prove that the fourth constraint is satisfied, we need to prove that the last constraint in (\ref{eq2:32}) is satisfied for $t\in[T]$, where $\beta_{u,T}$ is set as follows
    \begin{linenomath}
    \begin{align}
        \beta_{u,T}=&1-\gamma-\gamma\displaystyle{\sum_{l\in\mathcal{L}}}\displaystyle{\sum_{t'<T}}p'_{u,l,t'}q'_{u,l,t'}\text{Pr}\{d_l\geq T-t'+1\}-\gamma\displaystyle{\sum_{l\in\mathcal{L}}}\frac{\theta_l}{\Delta_u}\displaystyle{\sum_{t'<T}p'_{u,l,t'}(1-q'_{u,l,t'})}.
    \end{align}
    \end{linenomath}

    For $t=1$, we have $\beta_{u,1}=1-\gamma\geq0$ when $\gamma\in[0,\Delta_u/(3\Delta_u-1)]$. For $t=2$, we have $\beta_{u,2}=1-\gamma-(1-\sum_{l\in\mathcal{L}}A_{u,l,1})\gamma$, where
    \begin{linenomath}
    \begin{align}
        \sum_{l\in\mathcal{L}}A_{u,l,1}=&\sum_{l\in\mathcal{L}}p'_{u,l,1}q'_{u,l,1}\text{Pr}\{d_l=1\}+(1-\sum_{l\in\mathcal{L}}p'_{u,l,1})+\sum_{l\in\mathcal{L}}p'_{u,l,1}(1-q'_{u,l,1})\frac{\Delta_u-\theta_l}{\Delta_u}.
    \end{align}
    \end{linenomath}
    Because $q'_{u,l,1}\leq1$ and $\sum_{l\in\mathcal{L}}p'_{u,l,1}\leq1$, we have $\sum_{l\in\mathcal{L}}A_{u,l,1}\geq0$ and thus $\beta_{u,2}\geq1-2\gamma$. We then have $\beta_{u,2}\geq0$ when $\gamma\in[0,\Delta_u/(3\Delta_u-1)]$.

    For $t\in[3,T]$, we first define $\lambda_{u,t}$ as
    \begin{linenomath}
    \begin{align}
        \lambda_{u,t}\defeq&\sum_{l\in\mathcal{L}}\frac{1}{\Delta_u}\sum_{t'<t}p'_{u,l,t'}(1-q'_{u,l,t'})\theta_l+\sum_{l\in\mathcal{L}}\sum_{t'<t}p'_{u,l,t'}q'_{u,l,t'}\text{Pr}\{d_l\geq t-t'+1\},
    \end{align}
    \end{linenomath}
    and then we have $\beta_{u,t}=1-\gamma-\gamma\lambda_{u,t}$. With $p'_{u,l,t}=\sum_{e\in\ E_u}x^*_{e,l,t}$ and $p'_{u,l,t}q'_{u,l,t}=\sum_{e\in\ E_u}q_ex^*_{e,l,t}$, we have
    \begin{linenomath}
    \begin{align}
        \lambda_{u,t}=&\frac{1}{\Delta_u}\sum_{l\in\mathcal{L}}\sum_{t'<t}\sum_{e\in E_u}x^*_{e,l,t}\big[(1-q_e)\theta_l+\theta q_e\text{Pr}\{d_l\geq t-t'+1\}\big]\nonumber\\
        &+\frac{\Delta_u-\theta}{\Delta_u}\sum_{l\in\mathcal{L}}\sum_{t'<t}\sum_{e\in E_u}x^*_{e,l,t}\text{Pr}\{d_l\geq t-t'+1\}.
    \end{align}
    \end{linenomath}
    By (\ref{eqn:offst1}) and (\ref{eqn:offst2}), we have $\lambda_{u,t}\leq1+(\Delta_u-1)/\Delta_u$, for all $t\in[T]$. When $\beta_{u,t}=1-\gamma-\gamma\lambda_{u,t}\geq0$, we need $1-(\lambda_{u,t}+1)\gamma\geq0$. It is straightforward that $\lambda_{u,t}\geq0$, so the above condition is equivalent to $\gamma\in[0,\Delta_u/(3\Delta_u-1)]$. In other words, when $\gamma\in[0,\Delta_u/(3\Delta_u-1)]$, the last constraint in (\ref{eq2:32}) is satisfied for all $t\in[T]$.

    For the fourth constraint in (\ref{eq2:32}), we utilize the conclusion that $\beta_{u,T}\geq0$ when $\gamma\in[0,\Delta_u/(3\Delta_u-1)]$. Then by the analysis for the third constraint, the following inequality is valid:
    \begin{linenomath}
    \begin{align}
        \alpha_{u,T}+\beta_{u,T}\leq&\sum_{t'\in[T-2]}\alpha_{u,t'}\sum_{l\in\mathcal{L}}B_{u,l,t'}\text{Pr}\{d_l=T-t'\}+\beta_{u,T-1}+\alpha_{u,T-1}\sum_{l\in\mathcal{L}}A_{u,l,T-1}.
    \end{align}
    \end{linenomath}
    With $\beta_{u,T}\geq0$, we have $\alpha_{u,T}\leq\alpha_{u,T}+\beta_{u,T}$. The fourth constraint is satisfied when $\gamma\in[0,\Delta_u/(3\Delta_u-1)]$. $\alpha_{u,t}$ and $\beta_{u,t}$ defined in (\ref{eq2:33}) is a valid solution to (\ref{eq2:32}).

4) In the fourth step, we provide a valid lower bound for $\Tilde{R}^{\Delta_u}_{u,1}$ with $\alpha_{u,t}$ and $\beta_{u,t}$ defined in (\ref{eq2:33}). With $\gamma=\Delta_u/(3\Delta_u-1)$, $\alpha_{u,t}$ and $\beta_{u,t}$ defined in (\ref{eq2:33}) is a valid solution to (\ref{eq2:32}), and the result of the objective function for (\ref{eq2:32}) is
\begin{linenomath}
\begin{align}
    \frac{\Delta_u}{3\Delta_u-1}\sum_{t\in[T]}\sum_{l\in\mathcal{L}}B_{u,l,t}r'_{u,l,t}
    =\frac{\Delta_u}{3\Delta_u-1}\sum_{t\in[T]}\sum_{l\in\mathcal{L}}\sum_{e\in E_u}q_er_{u,v,l}x^*_{e,l,t}.\label{eq2:35}
\end{align}
\end{linenomath}
    
According to the linear programming duality theorem, we have
\begin{linenomath}
\begin{equation}
    \min\Tilde{R}^{\Delta_u}_{u,1}=\max\sum_{t\in[T]}\alpha_{u,t}\sum_{l\in\mathcal{L}}B_{u,l,t}r'_{u,l,t},
\end{equation}
\end{linenomath}
and
\begin{linenomath}
\begin{equation}
    \min\Tilde{R}^{\Delta_u}_{u,1}\geq\frac{\Delta_u}{3\Delta_u-1}\sum_{t\in[T]}\sum_{l\in\mathcal{L}}\sum_{e\in E_u}q_er_{u,v,l}x^*_{e,l,t}.
\end{equation}
\end{linenomath}
Therefore, Lemma \ref{lm:6} is proved.

\subsection{Proof of Lemma \ref{lm:7}}

By Lemma \ref{lm:1}, we have
\begin{linenomath}
\begin{equation}
    \text{LP(\texttt{Off})}=\displaystyle{\sum_{t\in[T]}\sum_{e\in E}q_e\sum_{l\in\mathcal{L}}}r_{u,v,l} x^*_{e,l,t}\geq\mathbb{E}_{I\sim \mathcal{I}}[\text{OPT}(I)].\label{eq5:73}
\end{equation}
\end{linenomath}
By Lemmas \ref{lm:22} and \ref{lm:6}, we have
\begin{linenomath}
\begin{equation}
    R^{\Delta_u}_{u,1}\geq\Tilde{R}^{\Delta_u}_{u,1}\geq\frac{\Delta_u}{3\Delta_u-1}\sum_{t\in[T]}\sum_{l\in\mathcal{L}}\sum_{e\in E_u}q_er_{u,v,l}x^*_{e,l,t}.\label{eq5:74}
\end{equation}
\end{linenomath}
By the definition of the activation value, the expected total reward of OMLA in the original system $\mathbb{E}_{I\sim\mathcal{I}}[\text{ALG}(I)]$ is equal to $\sum_{u\in U}R^{\Delta_u}_{u,1}$ (for the limited rejection case). Then by (\ref{eq5:73}) and (\ref{eq5:74}), we have
\begin{linenomath}
\begin{align}
    \frac{\mathbb{E}_{I\sim\mathcal{I}}[\text{ALG}(I)]}{\mathbb{E}_{I\sim\mathcal{I}}[\text{OPT}(I)]}\geq&\frac{\displaystyle{\sum_{u\in U}}R^{\Delta_u}_{u,1}}{\text{LP(\texttt{Off})}}\nonumber\\
    \geq&\frac{\displaystyle{\sum_{u\in U}\frac{\Delta_u}{3\Delta_u-1}\sum_{t\in[T]}\sum_{l\in\mathcal{L}}\sum_{e\in E_u}}q_er_{u,v,l}x^*_{e,l,t}}{\text{LP(\texttt{Off})}}.\label{eq5:75}
\end{align}
\end{linenomath}
Lemma \ref{lm:7} is proved.

\subsection{Proof of Theorem \ref{tr:2}}

By Lemma \ref{lm:7}, we have
    \begin{linenomath}
    \begin{align}
        \frac{\mathbb{E}_{I\sim\mathcal{I}}[\text{ALG}(I)]}{\mathbb{E}_{I\sim\mathcal{I}}[\text{OPT}(I)]}\geq&\frac{\displaystyle{\sum_{u\in U}\frac{\Delta_u}{3\Delta_u-1}\sum_{t\in[T]}\sum_{l\in\mathcal{L}}\sum_{e\in E_u}}q_er_{u,v,l}x^*_{e,l,t}}{\displaystyle{\sum_{u\in U}\sum_{t\in[T]}\sum_{l\in\mathcal{L}}\sum_{e\in E_u}}q_er_{u,v,l}x^*_{e,l,t}}\nonumber\\
        \geq&\min_{u\in U}\frac{\displaystyle{\frac{\Delta_u}{3\Delta_u-1}\sum_{t\in[T]}\sum_{l\in\mathcal{L}}\sum_{e\in E_u}}q_er_{u,v,l}x^*_{e,l,t}}{\displaystyle{\sum_{t\in[T]}\sum_{l\in\mathcal{L}}\sum_{e\in E_u}}q_er_{u,v,l}x^*_{e,l,t}}\nonumber\\
        =&\min_{u\in U}\frac{\Delta_u}{3\Delta_u-1}.
    \end{align}
    \end{linenomath}
    We define $\Delta=\max_{u\in U}\Delta_u$, then we have
    \begin{linenomath}
    \begin{align}
        \frac{\mathbb{E}_{I\sim\mathcal{I}}[\text{ALG}(I)]}{\mathbb{E}_{I\sim\mathcal{I}}[\text{OPT}(I)]}\geq&\min_{u\in U}\frac{\Delta_u}{3\Delta_u-1}\nonumber\\
        =&\frac{\Delta}{3\Delta-1}.\label{eq5:77}
    \end{align}
    \end{linenomath}
    By (\ref{eq5:77}), Theorem \ref{tr:2} is proved.

\end{document}